\documentclass[preprint]{aastex62}

\usepackage{amssymb}
\usepackage[caption=false]{subfig} 
\usepackage{float}

\shorttitle{White Dwarf Parallaxes}
\shortauthors{Leggett et al.}

\begin{document}


\title{Distant White Dwarfs in the US Naval Observatory Flagstaff Station Parallax Sample}


\author{S. K. Leggett}
\affiliation{Gemini Observatory, Northern Operations Center,  670 N. A'ohoku Place, Hilo, HI 96720, USA}
\affiliation{US Naval Observatory, Flagstaff Station, 10391 W. Naval Observatory Road, Flagstaff, AZ 86005$-$8521, USA}
\email{sleggett@gemini.edu}
\author{P. Bergeron}
\affiliation{D\'{e}partement de Physique, Universit\'{e} de Montr\'{e}al, C.P. 6128, Succ. Centre$-$Ville, Montr\'{e}al, Qu\'{e}bec H3C 3J7, Canada}
\author{John P. Subasavage}
\affiliation{US Naval Observatory, Flagstaff Station, 10391 W. Naval Observatory Road, Flagstaff, AZ 86005$-$8521, USA}
\affiliation{The Aerospace Corporation, 2310 E. El Segundo Boulevard, El Segundo, CA 90245, USA}
\author{Conard C. Dahn}
\affiliation{US Naval Observatory, Flagstaff Station, 10391 W. Naval Observatory Road, Flagstaff, AZ 86005$-$8521, USA}
\author{Hugh C. Harris}
\affiliation{US Naval Observatory, Flagstaff Station, 10391 W. Naval Observatory Road, Flagstaff, AZ 86005$-$8521, USA}
\author{Jeffrey A. Munn}
\affiliation{US Naval Observatory, Flagstaff Station, 10391 W. Naval Observatory Road, Flagstaff, AZ 86005$-$8521, USA}
\author{Harold D. Ables}
\affiliation{US Naval Observatory, Flagstaff Station, 10391 W. Naval Observatory Road, Flagstaff, AZ 86005$-$8521, USA}
\author{Blaise J. Canzian}
\affiliation{L$-$3 Communications/Brashear, 615 Epsilon Dr., Pittsburgh, PA 15238$-$2807, USA}
\author{Harry H. Guetter}
\affiliation{US Naval Observatory, Flagstaff Station, 10391 W. Naval Observatory Road, Flagstaff, AZ 86005$-$8521, USA}
\author{Arne H. Henden}
\affiliation{AAVSO, Cambridge, MA 02138, USA}
\author{Stephen E. Levine}
\affiliation{Lowell Observatory, 1400 W. Mars Hill Road, Flagstaff, AZ  86001$-$4499, USA}
\author{Christian B. Luginbuhl}
\affiliation{US Naval Observatory, Flagstaff Station, 10391 W. Naval Observatory Road, Flagstaff, AZ 86005$-$8521, USA}
\author{Alice B. Monet}
\affiliation{US Naval Observatory, Flagstaff Station, 10391 W. Naval Observatory Road, Flagstaff, AZ 86005$-$8521, USA}
\author{David G. Monet}
\affiliation{US Naval Observatory, Flagstaff Station, 10391 W. Naval Observatory Road, Flagstaff, AZ 86005$-$8521, USA}
\author{Jeffrey R. Pier}
\affiliation{US Naval Observatory, Flagstaff Station, 10391 W. Naval Observatory Road, Flagstaff, AZ 86005$-$8521, USA}
\author{Ronald C. Stone}
\altaffiliation{Deceased}
\affiliation{US Naval Observatory, Flagstaff Station, 10391 W. Naval Observatory Road, Flagstaff, AZ 86005$-$8521, USA}
\author{Frederick J. Vrba}
\affiliation{US Naval Observatory, Flagstaff Station, 10391 W. Naval Observatory Road, Flagstaff, AZ 86005$-$8521, USA}
\author{Richard L. Walker}
\altaffiliation{Deceased}
\affiliation{US Naval Observatory, Flagstaff Station, 10391 W. Naval Observatory Road, Flagstaff, AZ 86005$-$8521, USA}
\author{Trudy M. Tilleman}
\affiliation{US Naval Observatory, Flagstaff Station, 10391 W. Naval Observatory Road, Flagstaff, AZ 86005$-$8521, USA}
\author{Siyi Xu}
\affiliation{Gemini Observatory, Northern Operations Center,  670 N. A'ohoku Place, Hilo, HI 96720, USA}
\author{P. Dufour}
\affiliation{D\'{e}partement de Physique, Universit\'{e} de Montr\'{e}al, C.P. 6128, Succ. Centre$-$Ville, Montr\'{e}al, Qu\'{e}bec H3C 3J7, Canada}

\begin{abstract}

This paper presents new trigonometric parallaxes and proper motions for 214 stars. The measurements were made at the US Naval Observatory Flagstaff Station (NOFS)  between 1989 and 2017, and the average uncertainty in the parallax values is 0.6 mas. We find good agreement with {\it Gaia} Data Release 2 measurements for the stars in common, although there may be a small systematic offset similar to what has been found  by other investigators. The sample is matched to catalogs and  the literature to create a photometric dataset  which spans the ultraviolet to the mid-infrared. New mid-infrared photometry is obtained for nineteen stars from archived {\it Spitzer} mosaics. New optical spectroscopy is presented for seven systems and additional  spectra were obtained from  the literature.  We identify a sub-sample of 179 white dwarfs (WDs) at distances of 25 -- 200~pc. Their spectral energy distributions (SEDs) are analyzed using model atmospheres.  The models reproduce the entire flux-calibrated SED very well and provide the atmospheric chemical composition, temperature, surface gravity, mass and cooling age of each WD. Twenty-six WDs are newly classified and twelve systems are presented as candidate unresolved binaries. We confirm one WD$+$red dwarf system and identify two WDs as candidate dust disk systems.   Twelve old and high-velocity systems are identified as candidate thick disk or halo objects. The WDs in the sample generally have Galactic disk-like ages of $< 8$~Gyr and masses close to the canonical 0.6~$M_{\odot}$.
\end{abstract}

\vskip 0.5in
\section{Introduction}

White dwarfs (WDs) are the remnants of the vast majority of stars. These Earth-sized objects have degenerate cores and typically a thin H or He atmosphere; nuclear fusion no longer occurs and they cool slowly. WDs as cold as 3000~K are known; these left the main-sequence $\sim 10$~Gyrs ago and hence are old disk or halo objects \citep[e.g.][]{Gianninas2015}.   WDs can provide insights into Galactic populations and the history of star formation \citep[e.g.][]{Kilic2017}.  
Massive WDs and WDs in close binary systems constrain supernova  models \citep[e.g.][]{Kilic2014} and the
interiors of pulsating WDs shed light on details of stellar evolution \citep[e.g.][]{Giammichele2018}. Many WDs have atmospheres polluted by disrupted planetary systems \citep{Barber2016,Jura2014,Zuckerman2007}.
Hence an improved understanding of WDs can contribute to all major areas of astrophysical research: planetary systems, the stars, the Galaxy and other galaxies. 

Knowledge of a WD's mass, age and composition is key to extracting the important information contained in a WD. 
There are two common approaches to determining these parameters: the spectroscopic method which relies on absorption line profiles,
and the photometric method which relies on the luminosity and spectral energy distribution (SED) of the WD
\citep{Bergeron1992, Bergeron2001, Bergeron2011, Giammichele2012, Gianninas2011, Kilic2010, Limoges2015, Tremblay2009}.
The photometric method is the only way to measure atmospheric parameters of WDs that do not show absorption lines in
their spectra, for example WDs cooler than $\sim$ 5000 K.
In this paper we present new trigonometric parallaxes for more than 200 stars measured at the US Naval Observatory Flagstaff Station (NOFS).  The majority of the sample are WDs, and about one-third of the WDs have an effective temperature ($T_{\rm eff}$) $\lesssim 5000$~K.  
We use recent atmospheric models to analyse the WDs using the photometric method, for which the parallax must be well determined.
For some of the warmer WDs we check the viability of the photometric analysis by comparing synthetic and observed optical spectra.

For 52 years the NOFS has been determining trigonometric parallaxes of faint high proper motion stars. NOFS parallaxes for 309 late-type dwarfs and subdwarfs were recently published \citep{Dahn2017}, and new NOFS parallaxes are included in a recent paper studying the sample of WDs that lie within 25~pc of the Sun \citep{Subasavage2017}. This paper presents new NOFS trigonometric parallaxes of WDs  at distances of 25~pc to 1.2~kpc, as well as parallaxes measured for other objects of interest that are on the NOFS program and have not been previously published. 

Section 2 presents new astrometric data for 214 stars and Section 3 presents optical and infrared photometry for the sample. New optical spectra for seven WDs are presented in Section 4. In Section 5 we describe the models and the fitting technique used to determine the composition, temperature and surface gravity of each WD's atmosphere, from which mass and cooling age are derived. Section 6 gives a broad overview of the entire sample and Section 7 described the observational properties of 179 WDs in the sample. Section 8 discusses the physical properties of the 179 WDs. Our conclusions are given in Section 9. Supporting material is given in the Appendix.

%\clearpage
\section{Astrometry}

\subsection{New Astrometry}
 
Table 1 lists 216 new NOFS astrometric measurements of 214 stars. 
Two targets --- WD 0003$+$177 and WD 1042$+$593 --- had two independent determinations carried out using different
CCD cameras  to provide a consistency check. Note that in the rest of this paper we omit the letters ``WD'' in front of object names for clarity. 

\citet{Dahn2017} give the history of the NOFS parallax program, including a description of the different filters and cameras used.
Table 1 gives relative and absolute trigonometric parallaxes, proper motion, and the implied tangential velocity. The filters and cameras used for each measurement are also given.
The methods used to determine parallax and proper motion are the same as described in previous NOFS publications  
\citep{Monet1983}. In brief, image centroids are determined and solutions for parallax are executed independently in right ascension and declination. The final value is a weighted average of the individual measurements. 
For this sample, the uncertainty in the parallax ranges from 0.2~mas to 1.6~mas, with an average of 0.6~mas.

The correction from relative to absolute parallax is carried out using the procedures described in detail by \citet{Harris2016}. For six stars there is no correction to absolute parallax because 
these objects are in highly reddened regions where no reliable reference star distance estimates were available. The relative parallaxes will be lower limits; comparison to the {\it Gaia} Data Release 2 (DR2) parallax values \citep{Brown2018} suggest a correction to absolute of $+0.4 \pm 0.3$~mas for these six stars.

The proper motion of the target star with respect to the mean proper
motion of the reference stars is presented as the total relative proper
motion in Table 1.  As discussed in \citet{Dahn2017}, we choose
to present the relative total proper motions to 0.1~mas~yr$^{-1}$
and the position angle of total motion to  0.1~degree.
We do not attempt to convert these relative proper motions to absolute values
because, as noted by Dahn et al., the corrections are 2 -- 6~mas~yr$^{-1}$ \citep{Harris2016}
and the vast majority of the targets have a large proper motion.
On the other hand, the ten targets with measured relative proper motions 
$\lesssim 20$ ~mas~yr$^{-1}$
(0015$+$004, 0939$+$071, 0956$-$017, 1053$-$092, 1126$+$185, 1219$+$130, 1711$+$335, 2006$+$481, 2157$-$079, PG2300$+$166)
may have an absolute proper motion significantly different from the value given in Table 1.  The tangential velocities
in Table 1 are presented primarily to flag high-velocity stars.

%\clearpage
\subsection{Comparison to {\it Gaia} DR2 Parallaxes}

{\it Gaia} DR2 occurred on 2018 April 25. The \href{http://www.cosmos.esa.int/web/gaia/dr2}{DR2 overview} gives the number of sources with positions, parallaxes and proper motions as $1.3 \times 10^9$ and estimates that the survey is complete for stars with 
{\it Gaia} $G$ magnitudes (a broad filter covering 330 -- 1050 nm) between 12 and 17, although an estimated 20\% of stars with proper motions $> 0\farcs 6$ yr$^{-1}$ are missing. More information is given in the \href{http://www.cosmos.esa.int/archive/documentation/GDR2/index.html}{Gaia DR2 documentation} and \citet{Brown2018, Lindegren2018}. 

Although the stars in our sample are near the faint limit of DR2 and predominantly have high proper motion, 90\% have {\it Gaia} DR2 trigonometric parallaxes.   Appendix Table 19 gives the DR2 location, parallax, proper motion and $G$ magnitude for these 196 stars, as well as 
the values of the astrometric goodness of fit statistic (ASTROMETRIC\_GOF\_AL), the astrometric excess noise measurement (ASTROMETRIC\_EXCESS\_NOISE) and the significance of the noise measurement (ASTROMETRIC\_EXCESS\_NOISE\_SIG). There are 36 stars
(identified in Table 19) for which the NOFS and  {\it Gaia} parallax values differ by more than twice the combined uncertainty, which is larger than would be expected for a Gaussian error distribution. Eighteen of these 36 are flagged in DR2 as having a poor astrometric fit and significant excess noise (Table 19). It is also possible that the  {\it Gaia} and/or NOFS uncertainties are underestimated. 
\citet{Lindegren2018} describes the  {\it Gaia} parallax uncertainties as underestimated by 8 -- 12\% for fainter sources such as those in Table 19, and the discrepancy for three of the 18 discrepant (but apparently well-measured) sources becomes $< 2~\sigma$ if the NOFS uncertainties are also underestimated by 10 -- 20\%. Finally, nine of the remaining 18 are known, suspect or candidate binary systems (Sections 6 and 8); the {\it Gaia} data model does not include orbital binary motion, and unresolved systems are treated as point sources by both {\it Gaia} and NOFS, so the astrometry for these sources could be affected by unmodeled motions.  A remainder of 9 sources with discrepant measurements in a (well-measured) sample of 178 is consistent with a Gaussian distribution.

\begin{figure}[!b]
\begin{center}
\vskip -0.8in
\includegraphics[angle=-90,width=0.85\textwidth]{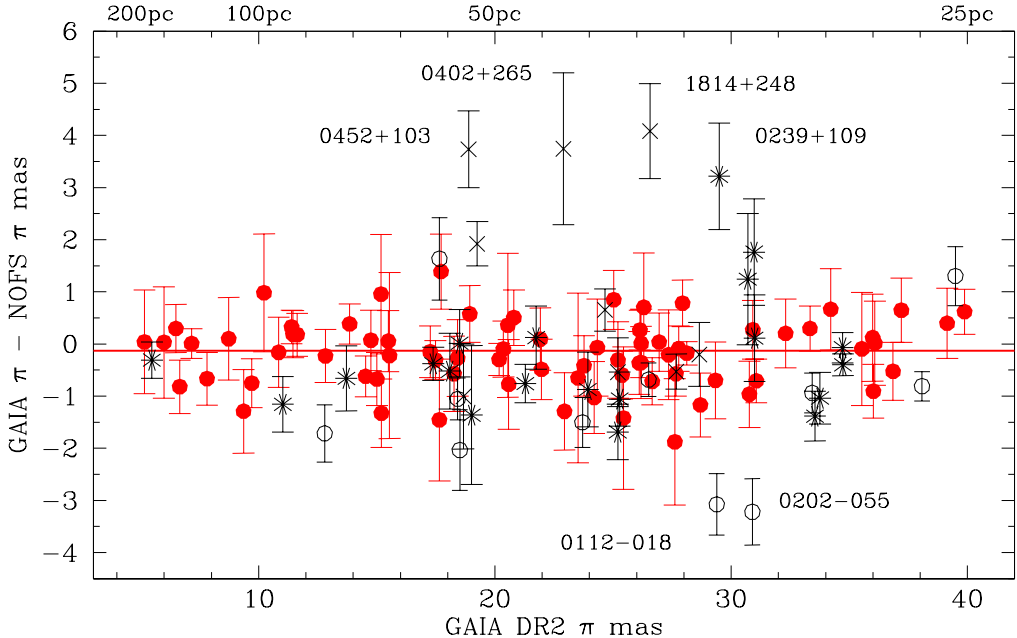}
\vskip -0.25in
\caption{The difference between the  absolute parallaxes from this work and {\it Gaia} Data Release 2 for 115 stars with  {\it Gaia} 
ASTROMETRIC\_GOF\_AL $< 3$ and ASTROMETRIC\_EXCESS\_NOISE\_SIG $< 2$. Asterisks indicate known binary systems and crosses indicate suspected binary systems. Open circles are stars for which the two measurements differ by $> 2~\sigma$. The red line is a weighted fit to the remaining 75 stars  (filled red circles); we find a   small 
negative offset in the {\it Gaia} DR2 parallaxes of $-130 \pm 62 ~\mu$as. 
}
%\vskip -0.2in
\end{center}
\end{figure}

Figure 1 shows the difference between the NOFS and {\it Gaia} DR2 parallax values for 115 stars in our sample which have absolute parallaxes
determined both by NOFS and  {\it Gaia}, and for which the  {\it Gaia} astrometric solution is robust as indicated by 
ASTROMETRIC\_GOF\_AL $< 3$ and ASTROMETRIC\_EXCESS\_NOISE\_SIG $< 2$ (see the \href{http://www.cosmos.esa.int/archive/documentation/GDR2/index.html}{Gaia DR2 documentation}). 
For these 115 stars the average uncertainty in the parallax value is 0.63~mas for NOFS measurements and 
0.15~mas for  {\it Gaia} measurements.
Systematic errors in the DR2 data are estimated to be below 0.1~mas and there is an average parallax offset of $\sim 30~\mu$as in the sense that the {\it Gaia}  DR2 parallaxes are too small \citep{Lindegren2018}.  Other work however indicates that the negative offset is 50 -- 110$~\mu$as \citep{Muraveva2018,Stassun2018}. 
In Figure 1 symbol types identify known and suspect binaries (Sections 6 and 8), and eleven notionally single stars for which the NOFS and  {\it Gaia} parallax values differ by more than $2~\sigma$. Similar to the discussion above, an increase of 10\% in the uncertainty of both measurements reduces the number of outliers to what would be expected for a Gaussian distribution.  Using the 75 sources which are not known or suspect binaries, and which have parallax measurements that agree within  $2~\sigma$ (red symbols in Figure 1), 
we find that the difference between the NOFS and {\it Gaia} parallaxes is independent of distance and a weighted fit gives a mean offset of
\begin{center} 
$ \pi (Gaia ~DR2) - \pi (NOFS) = -130 \pm 62 ~\mu$as 
\end{center}
with a spread around the mean of $260~\mu$as. 
The size of the negative offset is consistent with that found by \citet{Stassun2018}.

The six most discrepant stars are identified in Figure 1. Four of these have {\it Gaia} parallaxes $\sim 4$~mas larger than the NOFS value; one is a known double degenerate  \citep[0239$+$109,][]{Bergeron1990b}, another a suspected  double degenerate \citep[1814$+$248,][]{Rolland2015}, and we propose the other two are also double degenerate systems (0402$+$265, 0452$+$103; Sections 6 and 8). 

A comparison of the relative proper motions measured by NOFS to the absolute proper motion measured by {\it Gaia} shows the known double degenerate 0239$+$109 as the extreme outlier in this comparison also, with the {\it Gaia} proper motion being larger than the NOFS relative value by 21~mas~yr$^{-1}$. 
For this sample the average uncertainty in both the  {\it Gaia} absolute motion 
and the NOFS relative motion is 0.3~mas~yr$^{-1}$, with a correction to absolute for the NOFS values 
of $\pm$  2 -- 6 ~mas~yr$^{-1}$ depending on position in the sky \citep{Harris2016}. 
Using the trimmed sample of 75 sources, we find an rms scatter in the difference between the NOFS relative motion and the {\it Gaia}  absolute motion  of 11~mas~yr$^{-1}$, significantly smaller than the
average motion of the sample which is 470~mas~yr$^{-1}$.
%, and small enough that we do not explore the differences further.

%\clearpage
\section{Photometry}

The entire sample was matched to the optical photometric catalogs provided by the \href{http://www.sdss.org/dr14/}{Sloan  Digital Sky Survey (SDSS) Data Release 14} and the 
\href{https://panstarrs.stsci.edu/}{Pan-STARRS1 Data Release 1}.
The sample was also matched to the near-infrared photometric catalogs provided by 
\href{https://irsa.ipac.caltech.edu/Missions/2mass.html}{the Two Micron Sky Survey (2MASS)}, \href{http://wsa.roe.ac.uk/dr10plus_release.html}{the tenth Data Release of the UKIRT Infrared Deep Sky Survey (UKIDSS)},
\href{http://wsa.roe.ac.uk/uhsDR1.html}{the UKIRT Hemisphere Survey (UHS) J--Band Data Release},
 and the \href{http://horus.roe.ac.uk/vsa/theSurveys.html}{Visible and Infrared Survey Telescope for Astronomy (VISTA) public survey}. 
The near-infrared sky survey photometry was supplemented by values for individual objects taken from the literature 
\citep{BL02, Bergeron2005, Farihi2004, Hall2008, Harris1999, Kilic2010, Leggett1998}. 

Table 2 lists $ugrizyYJHK$ for the sample. 
%The $ugrizy$ are AB magnitudes and the $YJHK$ are Vega magnitudes. 
% Apermag3
The $u$ photometry is taken from the SDSS catalog, and  
$grizy$ are from the Pan-STARRS1 Data Release 1. The $r$ and $i$ filters are very similar for the two surveys, but $g$ and $z$ differ \citep{Chambers2016}. The Appendix Table 20 lists the complete set of SDSS photometry where available. Pan-STARRS photometry is used as the default here because it is 
available for all stars in our sample, includes an additional filter in the red, agrees well with the SDSS values for stars in common, and has similar or smaller uncertainties.
The $YJHK$ photometry in Table 2 is all on the MKO photometric system, as defined by the UKIDSS filters. The 2MASS $JHK_s$ magnitudes were converted to the UKIDSS system using transformations given in \citet{Hodgkin2009}. The VISTA $ZYJHK_s$ magnitudes were converted to the UKIDSS system using transformations given in 
\citet{Gonzalez-Fernandez2018}. Some of the additional literature values were given in the CIT/LCO system, and these were converted to MKO using  transformations given in \citet{Leggett2006}. We adopted the near-infrared magnitudes with the smaller uncertainties, or in the case of multiple values with similar uncertainties we adopted the average value. The Appendix Tables 21, 22, 23 and 24 list the 2MASS, UKIDSS, UHS and VISTA magnitudes in each of those photometric systems.

%\clearpage

%UNCOMMENT LINE FOR FULL TABLE
%\input{Table1}
%\startlongtable
\begin{longrotatetable}
\begin{deluxetable*}{lcccrrcrrr@{ $\pm$ }rr@{ $\pm$ }rr@{ $\pm$ }rr@{ $\pm$ }rr@{ $\pm$ }rrl}
%%\begin{splitdeluxetable*}{lllccrccrBr@{$\pm$}rr@{$\pm$}rr@{$\pm$}rr@{$\pm$}rr@{$\pm$}rrll}
\tabletypesize{\tiny}
\tablewidth{0pt}
\tablecaption{NOFS Astrometric Results
\label{tab:nofsast}}
\tablehead{
  \colhead{WD Number}                                 &
  \colhead{R.A.}                                 &
  \colhead{Decl.}                                 &
  \colhead{}                                 &
  \colhead{}                                 &
  \colhead{}                                 &
  \colhead{}                                 &
  \colhead{}                                 &
  \colhead{}                                 &
  \multicolumn{2}{c}{$\pi$(rel)}             &
  \multicolumn{2}{c}{$\pi$(corr)}            &
  \multicolumn{2}{c}{$\pi$(abs)}             &
  \multicolumn{2}{c}{$\mu$}                  &
  \multicolumn{2}{c}{P.A.}                   &
  \colhead{$V_{\rm tan}$}                    &
  \colhead{}                                 \\
  \colhead{or Name}                             &
  \colhead{J2000.0}                   &
  \colhead{J2000.0}                  &
  \colhead{Filter}                           &
  \colhead{$N_{\rm ngt}$}                    &
  \colhead{$N_{\rm frm}$}                    &
  \colhead{Coverage}                         &
  \colhead{Years}                            &
  \colhead{$N_{\rm ref}$}                    &
  \multicolumn{2}{c}{(mas)}                  &
  \multicolumn{2}{c}{(mas)}                  &
  \multicolumn{2}{c}{(mas)}                  &
  \multicolumn{2}{c}{(mas yr$^{\rm -1}$)}    &
  \multicolumn{2}{c}{(deg)}                  &
  \colhead{(km s$^{\rm -1}$)}                &
  \colhead{CCD}                              \\
  \colhead{(1)}                              &
  \colhead{(2)}                              &
  \colhead{(3)}                              &
  \colhead{(4)}                              &
  \colhead{(5)}                              &
  \colhead{(6)}                              &
  \colhead{(7)}                              &
  \colhead{(8)}                              &
  \colhead{(9)}                              &
  \multicolumn{2}{c}{(10)}                   &
  \multicolumn{2}{c}{(11)}                   &
  \multicolumn{2}{c}{(12)}                   &
  \multicolumn{2}{c}{(13)}                   &
  \multicolumn{2}{c}{(14)}                   &
  \colhead{(15)}                             &
  \colhead{(16)}                             
  }
\startdata
%              Name                                    RAJ2000          DEJ2000       Fil     Nni      Nfr         Coverage          Yrs      Nst     PI_rel   relerr    PIcor    coerr    PI_abs   aberr       PM      perr     Angle    A_err     Vtan      CCD     Notes
\object[LSR J0002$+$6357]{2359$+$636}         &  00:02:23.80  &  $+$63:57:44.1  &  A2-1  &  89   &  143  &  2008.57$-$2012.95  &  4.37   &  29  &  37.49  &  0.26  &  1.40  &  0.07  &  38.89  &  0.27 &   925.9  &  0.1  &   83.2  &  0.1  &   112.8 &  EEV24  \\
\object[WD 0002$+$729]{0002$+$729}            &  00:05:06.86  &  $+$73:13:09.4  &  A2-1  &  83   &  124  &  1998.57$-$2003.81  &  5.24   &  15  &  27.07  &  0.48  &  1.28  &  0.11  &  28.35  &  0.49 &   227.2  &  0.2  &   57.6  &  0.1  &    38.0 &  Tek2K  \\
\object[WD 0003$-$103]{0003$-$103}            &  00:05:55.91  &  $-$10:02:13.5  &  A2-1  &  97   &  112  &  2007.72$-$2014.82  &  7.10   &  11  &   5.35  &  0.41  &  0.85  &  0.06  &   6.20  &  0.41 &    82.9  &  0.1  &   67.4  &  0.1  &    63.4 &  Tek2K  \\
\object[WD 0003$+$177]{0003$+$177}            &  00:06:22.95  &  $+$18:00:13.4  &  ST-R  &  52   &  60   &  1990.78$-$1995.82  &  5.04   &  8   &  17.86  &  1.16  &  0.54  &  0.05  &  18.40  &  1.16 &   427.2  &  0.5  &  139.2  &  0.1  &   110.0 &  TI800  \\
\object[WD 0003$+$177]{0003$+$177}            &  00:06:22.95  &  $+$18:00:13.4  &  A2-1  &  72   &  78   &  1992.57$-$1996.95  &  4.39   &  10  &  16.37  &  0.53  &  0.83  &  0.07  &  17.20  &  0.53 &   419.4  &  0.4  &  139.1  &  0.1  &   115.5 &  Tek2K  \\
\enddata
\tablecomments{Table 1 is published in its entirety in the machine-readable format.  A portion is shown here for guidance regarding its form and content.}
\end{deluxetable*}
\end{longrotatetable}

%UNCOMMENT LINE FOR FULL TABLE  
%\input{Table2}
%\startlongtable
\begin{longrotatetable}
\begin{deluxetable}{lr@{ $\pm$ }rr@{ $\pm$ }rr@{ $\pm$ }rr@{ $\pm$ }rr@{ $\pm$ }rr@{ $\pm$ }rr@{ $\pm$ }rr@{ $\pm$ }rr@{ $\pm$ }rr@{ $\pm$ }r}
\tabletypesize{\tiny}
\tablewidth{0pt}
\tablecaption{Optical and Near-Infrared Photometry
\label{tab:phot}}
\tablehead{
\colhead{WD Number} & \multicolumn{2}{c}{$u$} &  \multicolumn{2}{c}{$g$} &  \multicolumn{2}{c}{$r$} &  \multicolumn{2}{c}{$i$} &
\multicolumn{2}{c}{$z$} &  \multicolumn{2}{c}{$y$} &  \multicolumn{2}{c}{$Y$} &  \multicolumn{2}{c}{$J$} &
 \multicolumn{2}{c}{$H$} &  \multicolumn{2}{c}{$K$} \\  
\colhead{or Name} &  \multicolumn{2}{c}{SDSS} &  \multicolumn{10}{c}{Pan-STARRS}   &  \multicolumn{8}{c}{MKO} 
}
\startdata
2359$+$636& \nodata  & \nodata  & 17.654  &  0.001 & 16.987  &  0.002 & 16.725 &   0.003 & 16.630 &   0.002 & 16.595 &   0.002& \nodata & \nodata & 15.784&0.067&15.561&0.136&15.516&0.216 \\
0002$+$729& \nodata  & \nodata  & 14.231  & 0.002 & 14.445 & 0.002&14.684&0.003&14.872&0.002&15.020&0.005& \nodata & \nodata & 14.614&0.036&14.568&0.055&14.757&0.100 \\
0003$-$103&17.346&0.014&17.746&0.005&18.199&0.007&18.524&0.013&18.796&0.012&18.962&0.025& \nodata & \nodata & \nodata & \nodata & \nodata & \nodata & \nodata & \nodata  \\
0003$+$177&17.412&0.029&17.053   &   0.005  & 16.981   &   0.001 &  17.046  &    0.003 &  17.133   &   0.004  & 17.192  &    0.013  & \nodata & \nodata & 16.208&0.126& \nodata & \nodata & \nodata & \nodata \\
0015$+$004&16.830&0.012&16.970&0.005&17.202&0.002&17.450&0.005&17.646&0.006&17.792&0.015&17.353&0.015&17.346&0.017&17.376&0.066&17.554&0.114 \\
\enddata 
\tablecomments{$ugrizy$ are AB magnitudes and $YJHK$ are Vega magnitudes. Magnitudes are Pan-STARRS MeanPSFMag,  and UKIDSS and VISTA apermag3.   Table 2 is published in its entirety in the machine-readable format. A portion is shown here for guidance regarding its form and content.}
\end{deluxetable}
\end{longrotatetable}

Mid-infrared photometry was also obtained for the sample from the \href{https://irsa.ipac.caltech.edu/Missions/wise.html}{Wide-field Infrared Survey Explorer (WISE) ALLWISE catalog}. {\it Spitzer} Space Telescope Infrared Array Camera (IRAC) photometry was also included for some stars ---  this camera has smaller pixels and is more sensitive than the {\it WISE} imager.  IRAC photometry was taken from the literature \citep{Barber2012,Barber2014,Barber2016,Bergfors2014,Farihi2008,Farihi2009,Farihi2010,Farihi2012,Jura2007,Kilic2009,
Kilic2010,Kilic2012,Mullally2007,Xu2012} and new photometry was also measured for this work using \href{https://irsa.ipac.caltech.edu/data/SPITZER/docs/spitzerdataarchives/}{processed images taken from the 
{\it Spitzer} data archive}. The {\it Spitzer} photometry was determined where the {\it WISE} data were missing or not consistent with the near-infrared photometry.
Table 3 lists the {\it Spitzer} fluxes measured here. Stars are identified in Table 3 for which the fluxes measured here differ significantly from previously published values using the same images. The agreement with the modelled fluxes support the values determined here. The stars with discrepant photometry are in crowded fields and care has to be taken with target and sky aperture placement;
possibly the more recent {\it Spitzer} pipeline produced cleaner images than were previously available. The Appendix Table 24 lists all {\it WISE} and  {\it Spitzer} photometry available, as Vega magnitudes.

\begin{deluxetable}{lclr@{ $\pm$ }rr@{ $\pm$ }rr@{ $\pm$ }rr@{ $\pm$ }r}
%\movetabledown=1.25in
%\def\arraystretch{0.8} 
%\tabletypesize{\small}
\tabletypesize{\scriptsize}
%\tablecolumns{9}
\tablewidth{0pt}
%\rotate
\tablecaption{New {\it Spitzer} Photometry}
\tablehead{
\colhead{WD Number} & \colhead{AOR} & \colhead{Principal} & 
\multicolumn{2}{c}{3.6 $\mu$m} &
 \multicolumn{2}{c}{4.5 $\mu$m} &
\multicolumn{2}{c}{5.8 $\mu$m} &
\multicolumn{2}{c}{8.0 $\mu$m}   \\
\colhead{or Name} & \colhead{Number} & \colhead{Investigator} &
\multicolumn{8}{c}{$\mu$Jy}
}
\startdata
0015$+$004 & 46975488 & Luhman & 30.30 & 0.15 & 21.12 & 0.13 & \nodata & \nodata & \nodata & \nodata \\
0235$+$064\tablenotemark{a} & 13106176 & Zuckerman & 207.52 & 6.35 & 122.76 & 5.34 & 70.88 & 4.73 & 21.84 & 2.73 \\
PG 0235$+$064B & 13106176 & Zuckerman & 16301.38 & 1.24 & 10827.47 & 1.01 & 9379.95 & 1.07 & 4648.51 & 3.22 \\
0507$+$045A & 10148352 & Kuchner & \nodata & \nodata & 107.11 & 0.68 & \nodata & \nodata & 61.43 & 1.37 \\
0507$+$045A & 39876352 & Luhman & \nodata & \nodata & 101.86 & 0.66 & \nodata & \nodata & \nodata & \nodata \\
0507$+$045B\tablenotemark{b} & 10148352 & Kuchner & \nodata & \nodata & 218.84 & 1.07 & \nodata & \nodata & 132.41 & 6.83 \\
0507$+$045B & 39876352 & Luhman & \nodata & \nodata & 208.16 & 4.27 & \nodata & \nodata & \nodata & \nodata \\
0919$+$296 & 17645568 & Fazio & 12.53 & 0.23 & 9.72 & 0.20 & \nodata & \nodata & \nodata & \nodata \\
1042$+$593 & 5175040,7770880 & Lonsdale & \nodata & \nodata & 38.43 & 1.07 & \nodata & \nodata & \nodata & \nodata \\
1235$+$422 & 54331904 & Trilling & \nodata & \nodata & 84.50 & 0.27 & \nodata & \nodata & \nodata & \nodata \\
1314$-$153 & 58372096 & Farihi & 210.16 & 0.95 & 145.18 & 1.07 & \nodata & \nodata & \nodata & \nodata \\
1439$-$195 & 61795584 & Trilling & 81.97 & 0.14 & 53.23 & 0.11 & \nodata & \nodata & \nodata & \nodata \\
1540$+$236 & 40159744,58045952 & Zuckerman & 227.49 & 0.23 & 143.13 & 0.18 & \nodata & \nodata & \nodata & \nodata \\
1552$+$177 & 47827456 & Jura & 54.08 & 0.20 & 37.21 & 0.17 & \nodata & \nodata & \nodata & \nodata \\
1845$+$019\tablenotemark{c} & 42671360 & Whitney & 7019.51 & 42.35 & 4910.5 & 4.27 & \nodata & \nodata & \nodata & \nodata \\
1912$+$143 & 45942528 & Whitney & 89.70 & 2.71 & 50.87 & 2.24 & \nodata & \nodata & \nodata & \nodata \\
2005$+$175 & 23614464 & Howell & \nodata & \nodata & 1405.32 & 1.05 & \nodata & \nodata & 1037.40 & 13.65 \\
2028$+$390 & 27107072,27107840 & Hora & 543.35 & 1.48 & 388.46 & 1.29 & 150.02 & 9.45 & \nodata & \nodata \\
2148$+$539 & 37871360 & Whitney & 74.11 & 5.29 & \nodata & \nodata & \nodata & \nodata & \nodata & \nodata 
\enddata 
\tablecomments{The absolute flux calibration uncertainty of 2\% \citep{Reach2005} is not included in the given errors.} \vskip -0.05in
\tablenotetext{a}{[3.6], [4.5] and [5.8] fluxes determined here are brighter than published by \citet{Farihi2008}.} \vskip -0.05in
\tablenotetext{b}{[8.0] flux  determined here is fainter than published by \citet{Mullally2007}.} \vskip -0.05in
\tablenotetext{c}{[3.6] and [4.5] fluxes  determined here are fainter than published by \citet{Barber2016}.} \vskip -0.05in
\vskip -0.3in
\end{deluxetable}

\clearpage
\section{New Optical Spectroscopy}

In this paper we explore the properties of the WDs primarily by comparing models to the SED as given by absolute photometric fluxes (Section 5). An optical spectrum can provide a check of the photometric fit or in some cases it can distinguish between  equally likely photometric solutions. For these reasons we obtained spectra of seven WDs, as well as the red dwarf companion to one of the WDs. Targets were selected which were accessible in the sky, which had no or poor-quality spectra available and where the spectrum would significantly contribute to the model analysis. Spectra covering 490 -- 820 nm were obtained at the Gemini North and South Observatories using the GMOS instruments \citep{Hook2004,Gimeno2016}. The data were obtained using Gemini's Fast Turnaround program, via programs GN-2018A-FT-209 and GS-2018A-FT-206.  The data were taken in thin cirrus with seeing that ranged from $0\farcs 5$ to $1\farcs 0$. 
The $1\farcs 0$ slit was used with the B600 grating, producing a resolution of 0.5~nm. Data were obtained for each star at wavelengths that differed by 10~nm so that chip gaps were covered. Flat fielding and wavelength calibration was done using calibration lamps mounted on the telescopes. The instrument sensitivity functions were determined using the calibration stars EG 131 at Gemini South and Feige 34 at Gemini North; final flux calibration was done using the Pan-STARRS $r$ and $i$ photometry for each target.  Figure 2 shows the new spectra for the seven WDs.
The observations are listed in Table 4, together with new spectral types for six of the seven targets.

\begin{figure}[!b]
\vskip -0.3in
\begin{center}
    \includegraphics[angle=0,width=0.6\textwidth]{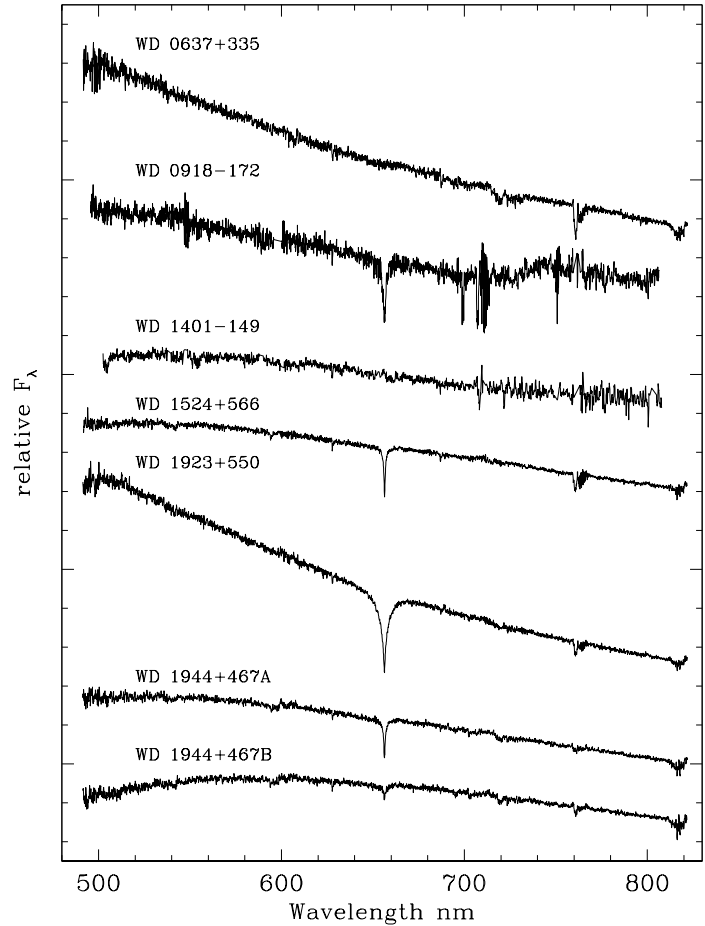}
\vskip -0.3in
\caption{New Gemini WD spectra obtained for this work. The spectrum of the dM6 companion to WD 0918$-$172 is not shown. The features near 760~nm are telluric. For WD 0918$-$172 difficulty determining the instrument sensitivity function at longer wavelengths likely produces a spurious rise in flux at $\lambda > 720$~nm.
}
\end{center}
\end{figure}
%\vskip -0.5in

\begin{deluxetable}{lcccccc}[!b]
%\movetabledown=1.25in
%\def\arraystretch{0.8} 
%\tabletypesize{\small}
\tabletypesize{\scriptsize}
%\tablecolumns{9}
\tablewidth{0pt}
%\rotate
\tablecaption{New Optical Spectroscopy}
\tablehead{
\colhead{WD Number} & \colhead{Coordinates} & \colhead{Instrument} & \colhead{Date} & \colhead{Exposure} & \multicolumn{2}{c}{Spectral Type} \\ 
\colhead{or Name}    &  \colhead{2018.3}  &  \colhead{} & \colhead{YYYYMMDD} & \colhead{minutes} & \colhead{Simbad}  & \colhead{ New}
}
\startdata
0637$+$335 & 06:40:33.59$+$33:27:34.6 & GMOS-N & 20180505, 20180506 & 18 & High proper motion star & DC \\
0918$-$172\tablenotemark{a} & 09:20:47.89$-$17:28:57.9 & GMOS-S & 20180419 & 18 & DA & DA\\
LP 787-25\tablenotemark{a} & 09:20:47.02$-$17:29:01.6 & GMOS-S & 20180419 & 18 & High proper motion star & dM6\tablenotemark{b} \\
1401$-$149 & 14:03:42.57$-$15:14:21.8 & GMOS-S & 20180421 & 36 &  High proper motion star & DC \\
1524$+$566 & 15:25:42.75$+$56:29:05.8 & GMOS-N & 20180523 & 16 & DC9 & DA \\
1923$+$550 & 19:24:09.68$+$55:06:52.1 & GMOS-N & 20180521 & 16 & dM7.5\tablenotemark{c} & DA \\
1944$+$467B & 19:45:21.42$+$46:50:01.7 & GMOS-N & 20180526 & 18 & High proper motion star & DA \\
1944$+$467A & 19:45:21.44$+$46:50:10.4  & GMOS-N & 20180526 & 18 &  High proper motion star & DA 
\enddata 
\tablenotetext{a}{0918$-$172 and LP 787-25 form a binary system with separation $13\farcs0$.}
\vskip -0.05in
\tablenotetext{b}{Typed by comparison to SDSS spectral templates \citep{Bochanski2007}.}
\vskip -0.05in
\tablenotetext{c}{Simbad gives the spectral type for the red dwarf companion determined by \citet{Kirkpatrick2010}.}
\vskip -0.3in
\end{deluxetable}

%\clearpage
\section{Model Atmospheres and Fitting Technique}

The analyses presented here  involve fitting the observed flux-calibrated WD SEDs with
pure-H and pure-He atmospheric models using a least-squares method; models are also available for more unusual compositions. \citet{Holberg2006} provide more information on the flux calibration process. 
The models for the hydrogen-atmosphere WDs
are built from the code described in \citet{Bergeron1995} and references therein, with recent improvements discussed in \citet{Tremblay2009}. The helium-atmosphere models are described in \citet{Bergeron2011}.  Cooler models with mixed hydrogen-helium atmospheres are also available, as described in \citet{Gianninas2015}. The analyses of DQ (carbon-rich) and DZ (metal-rich) white dwarfs rely on the LTE model atmosphere calculations developed by \citet{Dufour2005,Dufour2007,Dufour2007Nature,Dufour2008}.

The fits are iterated allowing $T_{\rm eff}$, 
the solid angle $\pi (R/D)^2$ (where $R$ is stellar radius and $D$ is distance) and surface gravity  $g$ to vary. 
The model flux is interpolated at  $T_{\rm eff}$ and log~$g$ where log~$g$ is obtained from $R$ and evolutionary models similar to those described in \citet{Fontaine2001} but with C/O cores, $q({\rm He})\equiv \log M_{\rm He}/M_{\star}=10^{-2}$ and $q({\rm H})=10^{-4}$, which are representative of hydrogen-atmosphere white dwarfs, and $q({\rm He})=10^{-2}$ and $q({\rm H})=10^{-10}$, which are representative of helium-atmosphere white dwarfs. For the (few) WDs with mass $< 0.2~M_{\odot}$ the \citet{Althaus2013} mass-radius relationship  was used. 
Iterations end when the value of $R$ required for flux scaling the model is consistent with the value implied by $g$. 
The WD cooling age (the time since the star left the main-sequence) is determined once  $T_{\rm eff}$, mass and atmospheric composition are known.
The method is described in detail by \citet{BRL97}.

The atmospheric parameters can be further checked by comparing synthetic spectra to any observed line profiles, the depth and width of which are sensitive to $T_{\rm eff}$, $g$ and atmospheric composition \citep[e.g.][]{Limoges2015}. 
Spectra were obtained from the literature 
\citep{Bergeron1992,Bergeron2001,Bergeron2011,Giammichele2012,Gianninas2011,Greenstein1986,Liebert2005, 
Limoges2015,Napiwotzki2003,Oppenheimer2001,Reid2005,Rolland2018,Subasavage2007,Subasavage2008,Subasavage2009}
and from the 
\href{http://www.http://skyserver.sdss.org/dr14/en/tools/search/SQS.aspx}{SDSS spectroscopy archive}. 
New spectra were also obtained for this work, as described in Section 4.

Figure 3 shows examples of fits to a warm DB and a warm DA WD, as well as a cool WD which is too cool to show H$\alpha$ but for which the photometric SED constrains the atmosphere to be pure hydrogen. 
This Figure shows that the $u$-band is a powerful composition diagnostic, however it is excluded in the fitting process because its inclusion distorts the helium-rich fit. Figure 4 shows examples of fits to a DQ and a DZ WD.
The WDs in Figures 3 and 4 have photometry which covers a broad wavelength range. Usually very little flux emerges from the WD atmospheres at mid-infrared wavelengths, and those data are not used in the model fitting, however the Figures show that there is good agreement between the models and observations across the entire SED, including the mid-infrared.

%\clearpage
%\vskip -0.2in
%\begin{center}
\begin{figure}[b]
\vskip -0.3in
\hskip 0.2in
    \includegraphics[angle=-90,width=1.0\textwidth]{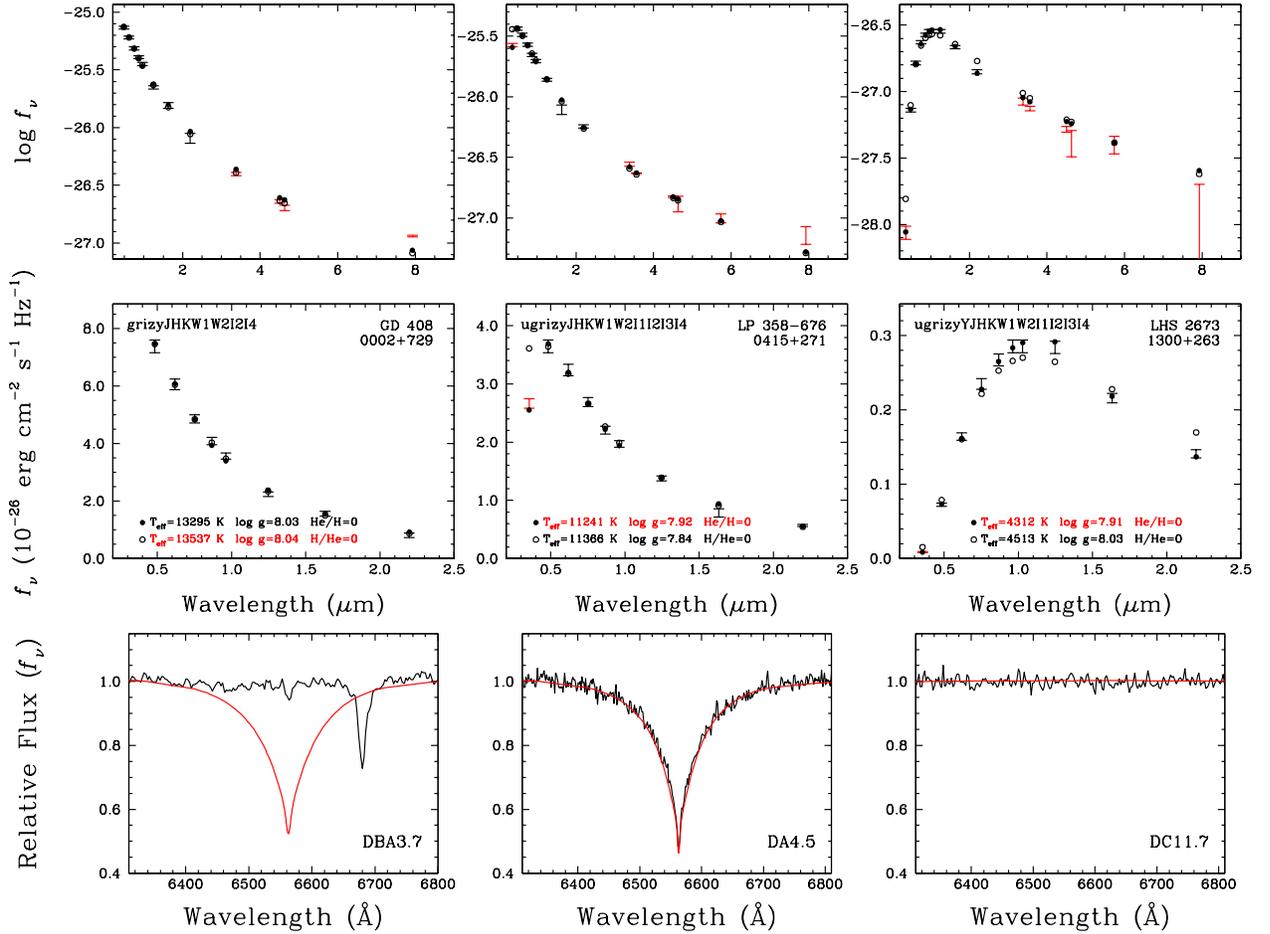}
\vskip -0.1in
 \caption{Examples of model fits to (left to right) a warm DB, a warm DA and a cool hydrogen-rich WD where the 
 SED constrains the atmospheric composition. In the upper plots error bars represent the observed fluxes through various filters and circles are the fitted model datapoints. Red error bars indicate that the bandpass was omitted from the fit. In the bottom panel the observed spectrum is in black and the hydrogen-rich model spectrum in red. The adopted model parameters are given in red in the middle panels.
}
\end{figure}
%\end{center}

%\vskip -0.8in
\begin{figure}[t]
\vskip -0.6in
    \includegraphics[angle=0,width=0.57\textwidth]{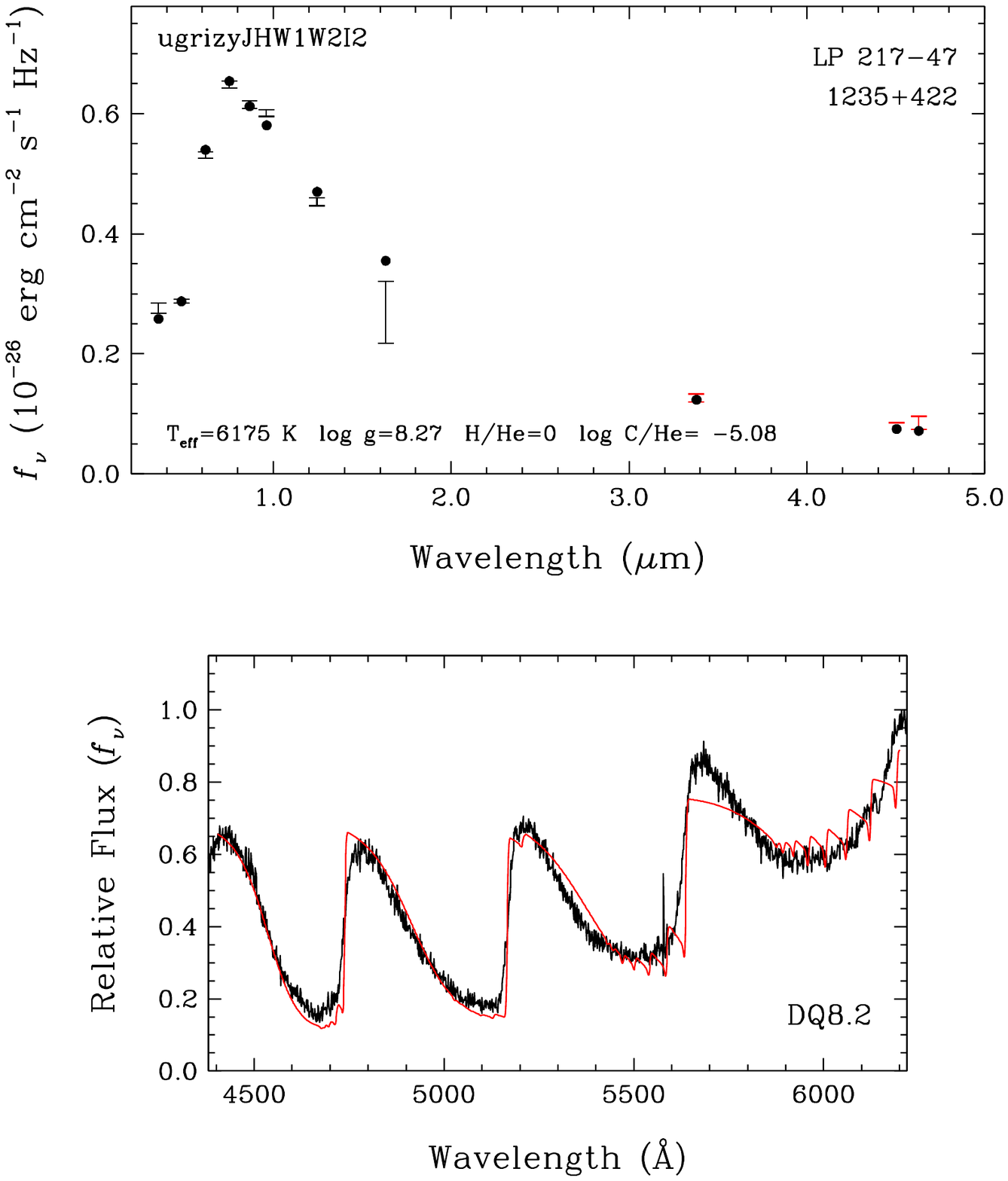}\\
\vskip -5.4in
\hskip 3.3in
    \includegraphics[angle=0,width=0.57\textwidth]{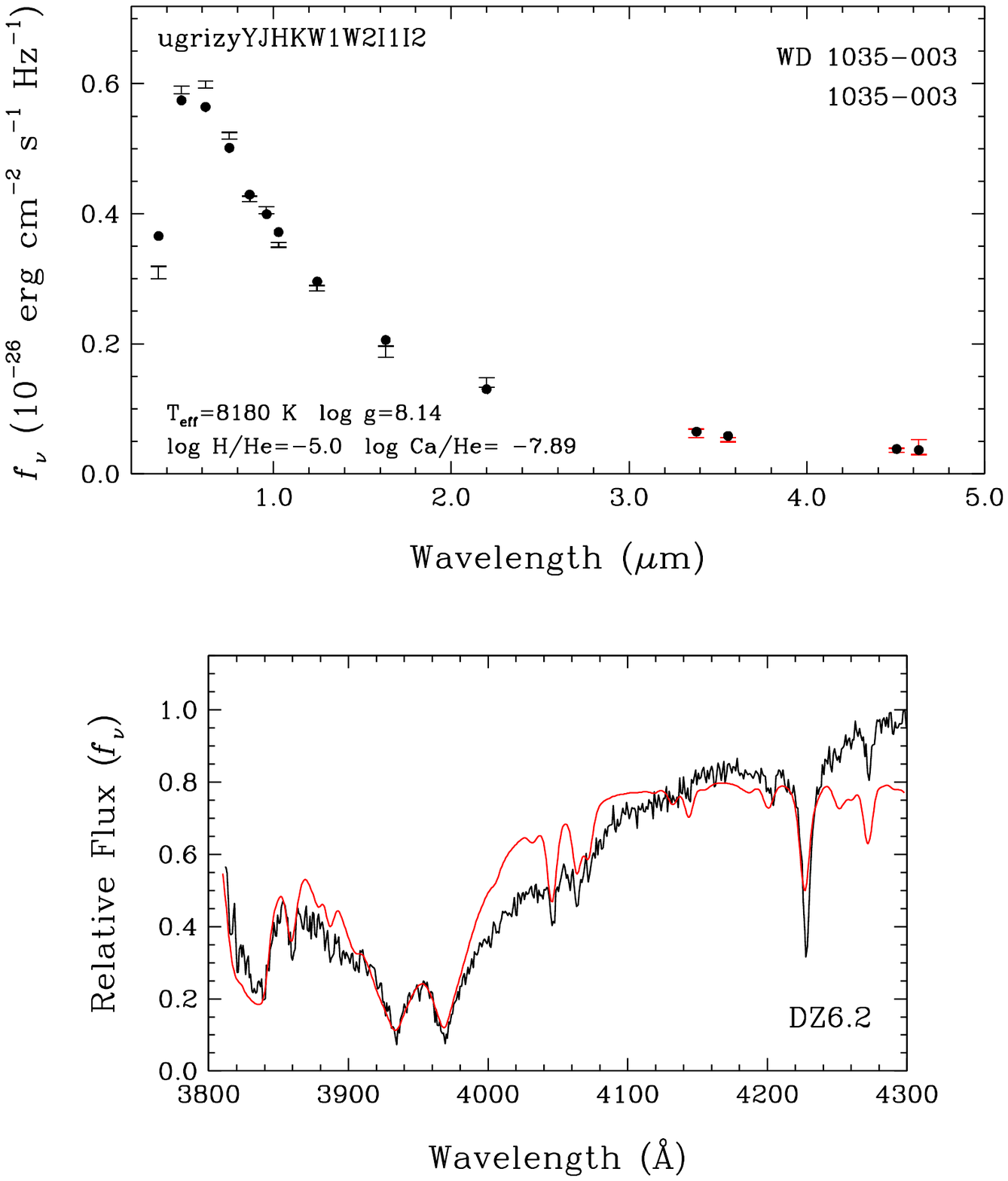}
\vskip -0.85in
\caption{Model fits to a DQ and a DZ. Error bars represent the observed fluxes and circles are the model datapoints. Red error bars indicate that the bandpass was omitted from the fit. The observed spectrum is shown in black; the red spectrum is  generated using the model parameters given in the top panels.
}
\end{figure}

\begin{deluxetable}{llll}
%\movetabledown=1.25in
%\def\arraystretch{0.8} 
%\tabletypesize{\small}
\tabletypesize{\scriptsize}
%\tablecolumns{9}
\tablewidth{0pt}
%\rotate
\tablecaption{Stars Not Analysed}
\tablehead{
\colhead{WD Number}   & \colhead{Name}   & \colhead{Simbad Spectral Type} & \colhead{Note, Revised Classification} 
}
\startdata
0104$+$509 & GD 274 & high PM star & subdwarf$+$K star \citep{Ulla1998} \\
0136$-$005 & LP 588-25 & M6V eclipsing binary & D:$+$dM \citep{Parsons2012} \\
 & PG 0235$+$064B & Infra-Red source & dM3.5 from SDSS spectrum and classification (this work)\\
0330$-$090 & V*  LL Eri  & DA$+$M2.5V &  DA$+$M2.5V \citep{Kawka2002}\\
 & LP 787-25 &  high PM star  & M dwarf (this work; Figures 5, 6) \\
 & G117-B15B & M3.5V & M3.5V  \citep{Kirkpatrick2011} \\
 & LHS 2140 & sdM0.5 &  sdM0.5 \citep{Gizis1997} \\
0939$+$071 & EGGR 431 &   DC7 & dF \citep{Gianninas2011} \\
1126$+$185 & PG 1126$+$186 & DC$+$G/K(e) & sdB$+$G/K \citep{Farihi2005}\\
1135$+$036 & V$^*$ T Leo &  dwarf nova & ultrashort period red dwarf + WD \citep{Shafter1984}  \\
1148$+$544 & LP 129-586 & DA5 & dM5 from SDSS spectrum and classification  (this work)\\
1232$+$379 & V* AM CVn &  cataclysmic variable & interacting DB WDs \citep{Ulla1990} \\
1303$+$182 & V* GP Com &  nova-like star & interacting DB WDs \citep{Ulla1990} \\
1711$+$335 & V* V795 Her &  nova & SW Sex star --- extreme mass transfer \citep{Schmidtobreick2017} \\
 & V$^*$ AM Her &  cataclysmic variable &  interacting magnetic WD and M4V \citep{Kawka2005} \\
2005$+$175 & V$^*$ WZ Sge &   dwarf nova & interacting WD and red or brown dwarf \citep{Kato2015}\\
2006$+$481 & & DB & sdOB \citep{Bergeron2000} \\
2154$+$408 & & DA1.7 & DA$+$dM3.5 \citep{Hillwig2002} \\
 & LP 400-21 & high PM star  &  M dwarf (this work; Figures 5, 6)  \\
2300$+$165 & PG 2300$+$166 & variable star &  subdwarf or subdwarf binary (this work; Figures 5, 6)  
\enddata 
\end{deluxetable}

%\clearpage
\vskip 1.0in
\section{Overview of the Sample}

\subsection{Core Sample of 179 Notionally Single WDs}

Figure 5 shows our entire parallax sample of 214 stars in a color-magnitude diagram which uses  Pan-STARRS photometry. Sequences for 0.6~$M_{\odot}$ WDs are shown, generated by the models described in Section 5. A 1~Gyr isochrone for solar-metallicity stars is also shown generated in Pan-STARRS colors using the \href{http://stev.oapd.inaf.it/cgi-bin/cmd}{PARSEC color-magnitude diagram web interface} \citep{Bressan2012,Marigo2013,Marigo2017,Rosenfield2016}. The majority of the sample --- 179 stars, or 84\% --- are single WDs, or have not been confirmed to be multiple and can be fit with our models as single WDs. 
%We show later that these WDs have 3000 $\lesssim T_{\rm eff}$~K $\lesssim$ 40000, and range in age from 0.1~Gyr to $\sim$15~Gyr. 
These 179 WDs form our core sample and the rest of this paper focusses on these stars.

\begin{figure}[b]
\vskip -1.5in
\begin{center}
    \includegraphics[angle=0,width=0.83\textwidth]{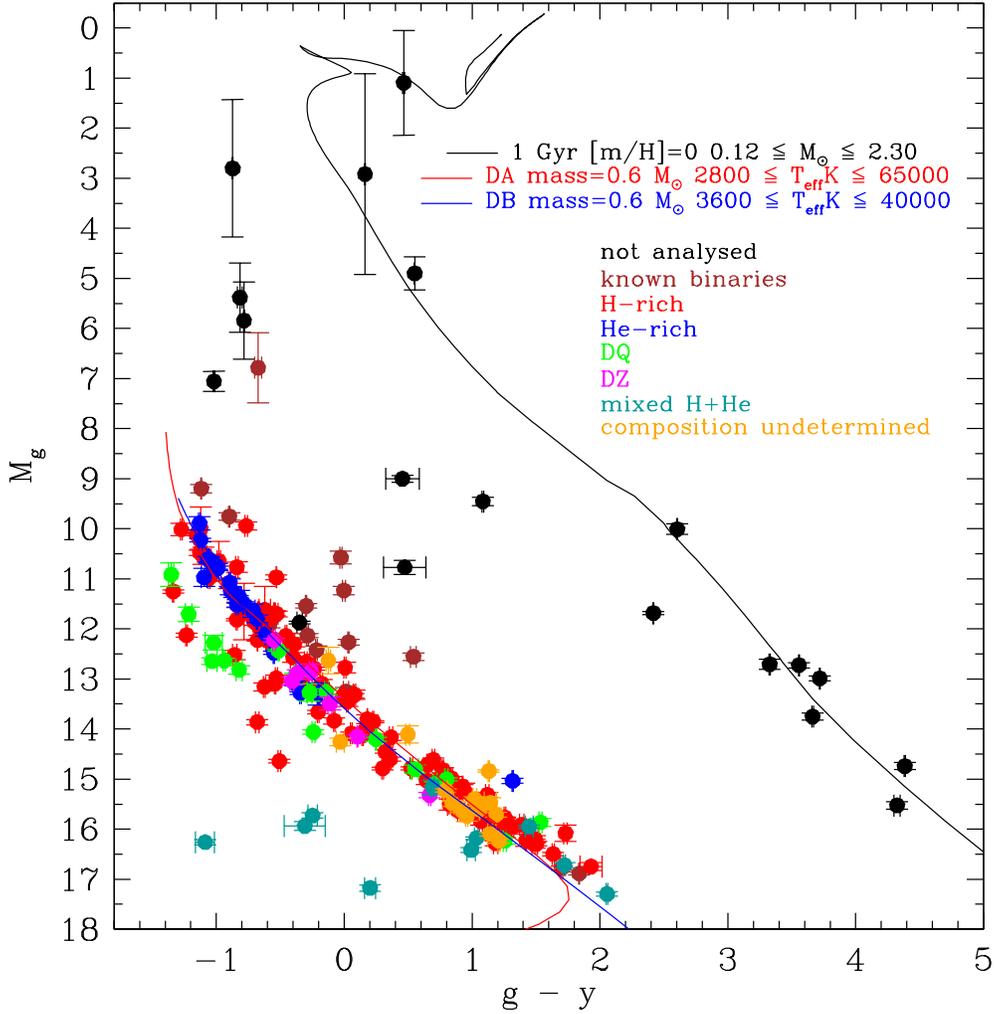}\\
\vskip -0.5in
 \caption{The $g-y:M_g$ color-magnitude diagram for the  214 stars in our sample. 
}
\end{center}
\end{figure}

\begin{figure}
\begin{center}
    \includegraphics[angle=-90,width=0.9\textwidth]{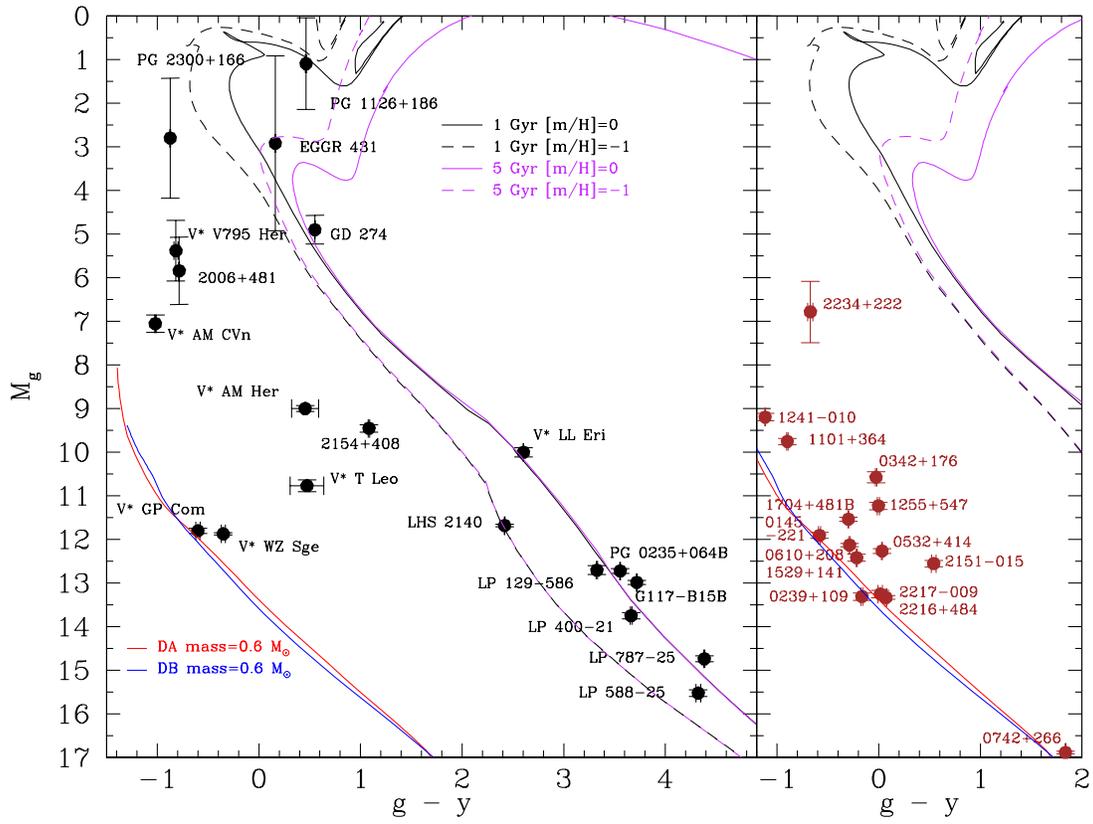}\\
\vskip -0.3in
 \caption{Same as Figure 5, for stars not analyzed (left panel) and unresolved binaries (right panel). 
}
\end{center}
\end{figure}

\begin{figure}[b]
\vskip -0.5in
\begin{center}
    \includegraphics[angle=-90,width=0.7\textwidth]{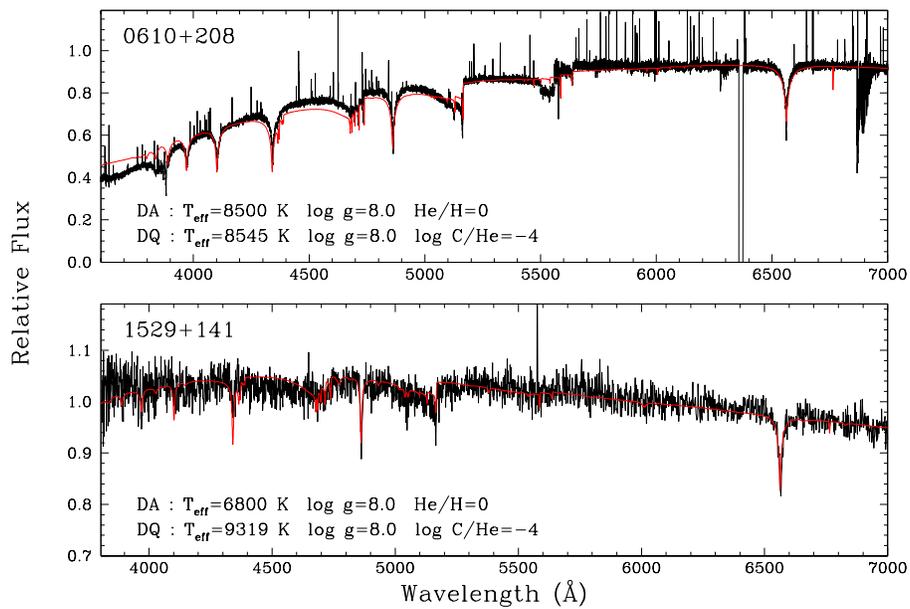}\\
\vskip -0.05in
 \caption{Preliminary fit to two DQ$+$DA binaries in the sample, see Table 6.
}
\end{center}
\end{figure}

\subsection{Other Stars in the Sample}

The remaining 35 stars are more complex. 
Twenty  
%(9\% of the full sample) 
are either not WDs or are WDs in systems too complex for this study. These 20 stars are listed in Table  5 and
identified in the left panel of Figure 6. Note that six stars with WD numbers are not in fact WDs: 
0104$+$509, 0939$+$071, 1126$+$185, 1148$+$544, 2006$+$481, and 2300$+$165. Three stars in Table 5 are newly classified as red dwarfs: 
PG 0235$+$064B, 1148$+$544 (LP 129-586), and LP 400-21.

Table 6 and the right panel of Figure 6 identify 
fifteen 
%(7\% of the full sample) 
known unresolved WD$+$WD or WD$+$dM/L binaries.  In some cases we can produce model fits   --- Figure 7 shows a preliminary deconvolution of the spectra of two DA$+$DQ binaries for example. However these fits are not rigorous and detailed analyses are postponed to a future paper. 
%The estimated properties of the WDs in the unresolved binary sample are given in Table 6.

%\setlength{\tabcolsep}{3pt}
\begin{deluxetable}{lllll}
%\movetabledown=1.25in
%\def\arraystretch{0.8} 
%\tabletypesize{\small}
\tabletypesize{\scriptsize}
%\tablecolumns{9}
\tablewidth{0pt}
%\rotate
\tablecaption{White Dwarfs in Unresolved Binaries}
\tablehead{
\colhead{WD Number}   & \colhead{Name}   & \colhead{Spectral Type} & \colhead{Binary Reference} &  \colhead{Model Fit}
}
\startdata
0145$-$221 & GD 1400 & DA4.6$+$dL &  \citet{Wachter2003} & dL optical flux contribution is negligible, DA  fit:\\
           &         &            &                     & $T_{\rm eff}=10959$~K, log $g=8.04$\\
0239$+$109 & G4-34 & DA$+$DC & \citet{Bergeron1990b}  & \citet{Gianninas2011} fit the spectrum and SED:\\
           &       &         &                       & $T_{\rm eff}=10060$~K DA and a  $T_{\rm eff}=7620$~K DC\\
0342$+$176 & LP 413-40 &  DA$+$DA &\citet{Kilic2010} & assuming the system is two identical DA WDs:\\
           &         &            &                     & $T_{\rm eff}=8314$~K, log $g=6.82$\\
0532$+$414 & GD 69 & D:$+$D: & \citet{Toonen2017} & assuming the system is a single DA:\\
           &       &       &                    & $T_{\rm eff}=7378$~K, log $g=7.14$\\
0610$+$208 & GD 73 & DQ$+$DA & \citet{Vennes2012} & Figure 7, the spectrum and SED can be fit as:\\
           &       &         &                      & $T_{\rm eff}=8500$~K DA and a  $T_{\rm eff}=8545$~K DQ\\
0742$+$266 &  LSPM J0745$+$2627 & D:$+$dM2 & \citet{Parsons2013} & dM flux significant, if fit as single WD:\\
          &       &         &                      & $T_{\rm eff}=3933$~K, log $g=7.92$ DC\\
1101$+$364 &        & D:$+$D: & \citet{Maxted1999} & assuming the system is a single DA:\\
         &       &         &                      & $T_{\rm eff}=13625$~K, log $g=6.92$\\
1241$-$010 &  PG  1241$-$010 & DA+DA+dM & \citet{Farihi2005} & dM flux contribution small, fit as single DA:\\
           &         &            &                     & $T_{\rm eff}=22732$~K, log $g=7.24$\\
1255$+$547 &  &  DA$+$DA & \citet{Marsh2011} & assuming the system is two identical DA WDs:\\
           &         &            &                     & $T_{\rm eff}=7414$~K, log $g=7.09$\\
1529$+$141 & GD 184 & DQ$+$DA & \citet{Giammichele2012} & Figure 7, the spectrum and SED can be fit as:\\
           &        &         &                & $T_{\rm eff}=6800$~K DA and a  $T_{\rm eff}=9319$~K DQ\\ 
1704$+$481B & EGGR 577 &   DA$+$DA & \citet{Maxted2000} & assuming the system is  a single DA:\\
          &         &            &                     & $T_{\rm eff}=8767$~K, log $g=7.18$ \\
2151$-$015 & EGGR 151 & DA5.6$+$dM8 & \citet{Maxted1999} & dM optical flux contribution is negligible, DA fit:\\
           &         &            &                     & $T_{\rm eff}=8989$~K, log $=7.98$\\
2216$+$484 & GD 402 & DC$+$DA & \citet{Bergeron1990b}  & \citet{Bergeron1990b} fit the spectrum and SED:\\
           &        &         &                       & $T_{\rm eff}=6200$~K DA and a  $T_{\rm eff}=7080$~K DC\\
2217$-$009 & PHL 5038 & DA6.8$+$dL8 &\citet{Steele2009} & dL optical flux contribution is negligible, DA fit:\\
           &         &            &                     & $T_{\rm eff}=7466$~K, log $g=7.95$\\
2234$+$222 & LP 400-22 & DA$+$DA & \citet{Kilic2009} & assuming the system  two identical DA WDs:\\
           &         &            &                     & $T_{\rm eff}=10716$~K, log $g=5.64$\\
\enddata 
\end{deluxetable}

\subsection{Interstellar Extinction and Reddening}

The core sample of 179 WDs includes stars at large distances. We corrected for reddening the observed fluxes for the 13 WDs  at $>$ 100~pc (Table 7), before fitting the  models. We use the approach described in \citet{Harris2006} except that we use  extinction vectors from \citet{Green2018}. Green et al. give vectors for Pan-STARRS and 2MASS filters, but not for MKO filters. We estimated the  MKO $YJHK$ extinction by applying the $\lambda^{-1}$ extinction dependence seen in the near-infrared \citep{Cardelli1989} to Pan-STARRS $y$ and 2MASS $JHK_s$ values.
%Table 7 lists the 13 WDs which we have corrected for reddening. 
%For each WD distance and Galactic latitude $b$ are given, as well as an indication of the impact of the correction on their derived parameters. 
The correction  makes the WD brighter --- $T_{\rm eff}$ increases and cooling age decreases. For the 11 WDs 
with $b > 10^{\circ}$,  $T_{\rm eff}$ increases by 4\%  and cooling age decreases by 10\%, on average. The changes to log $g$ and mass are small.
The object with the largest correction is  1910$+$047; the nominal correction produces unphysical  parameters and instead we applied 10\% of the maximum extinction which gives a $T_{\rm eff}$ value consistent with that determined by \cite{Vennes1990}
from a fit to the observed Balmer lines and 
gives values for $g$ and mass close to the canonical WD values. 
%\citep{Gianninas2011,Bergeron2011}.

\begin{deluxetable}{rcrrrrrrrrrr}[!t]
%\tabletypesize{\small}
\tabletypesize{\scriptsize}
%\tablecolumns{9}
\tablewidth{0pt}
\tablecaption{Reddening Correction for Distant White Dwarfs}
\tablehead{
\colhead{WD}   & \colhead{Spectral} &
\colhead{Distance}   & \colhead{Galactic} & \colhead{$T_{\rm eff}$} &  \colhead{Mass} &   \colhead{log $g$} & 
\colhead{Cooling} & \multicolumn{4}{c}{Corrected $-$ Uncorrected} \\
\colhead{Number}   & \colhead{Type} &
\colhead{pc}   & \colhead{Latitude deg.} & \colhead{K} &  \colhead{$M_{\odot}$} &  \colhead{dex}   &
\colhead{Age Myr} &
\colhead{$\delta(T_{\rm eff}$)} &  \colhead{$\delta$(log$g$)} &  \colhead{$\delta$(mass)} & \colhead{$\delta$(Age)} 
}
\startdata
0003$-$103   &    DQhot2.4 &  161     &        $-$69.8  &  20882  &  1.125 &   8.863  &  333   &      593  &    0.015  &  0.008  &  $-$19 \\
0015$+$004   &     DBP3.5 &  132    &         $-$61.0 &   14532 &   0.621 &   8.054 &   240  &       172 &     0.011 &   0.007  &  $-$4 \\
0112$+$104    &    DB1.8 &  116    &         $-$51.8 &   27942 &   0.518 &   7.818 &   13  &       582 &     0.014 &   0.007 &   $-$1 \\
0500$+$573   &   D:6.0 &    108    &         9.7   &   8226 &   0.426  &  7.700 &   706    &     379  &    0.080  &  0.039 &   $-$29 \\
0919$+$296  &    DQ6.2  &    116   &          44.3  &   8068  &    0.652 &  8.122 & 1277 & 19 & 0.002 & 0.001 & $-$4 \\
0954$+$342   &      DB1.9 &  207    &         52.3  &   27023 &   0.909 &   8.488 &   72    &     3760  &   0.057  &  0.039 &   $-$38 \\
0956$-$017   &   DAZ3.7 &    196    &         39.3  &   13526  &  0.674 &   8.108 &   315    &     1080 &     0.028 &   0.019 &   $-$64 \\
1002$+$430  &   DA2.5 &    124     &        53.1  &   19895  &  0.584 &   7.928  &  58    &     772  &    0.044  &  0.026 &   $-$4 \\
1053$-$092   &   DA2.2 &   195     &        43.9  &   23298 &   0.498  &  7.736  &  22     &    1592 &    0.063  &  0.032 &   $-$7 \\
1219$+$130   &    DAZ4.2 &   211     &        74.0  &   12018 &   0.573  &  7.943  &  345     &    603  &    0.026  &  0.016 &   $-$39 \\
1327$+$594   &  DQAPhot2.7 &     134     &        57.3  &   18755 &   1.194  &  9.011   & 640     &    655  &    0.031  &  0.014  &  $-$30 \\
1910$+$047   &    DA2.1 &    168     &        $-$2.4  &   24208 &   0.673 &   8.070 &   32   &      7540  &   0.300  &  0.179 &   $-$60 \\
2157$-$079   &   DQhot1.9 &    236     &        $-$45.1 &   26142 &   1.009 &   8.649 &   120   &      3363  &   0.109  &  0.070 &   $-$28 \\
\enddata 
\end{deluxetable}

\vskip 0.5in
%\clearpage
\section{Observational Properties of the Sample of 179 White Dwarfs}

\subsection{Atmospheric Composition and Trends with Color}

Table 8 lists the 179 WDs in this sample that are assumed to be single stars and which we have analyzed with the models described in Section 5. All the WDs in Table 8 have a numerical type based on our assigned $T_{\rm eff}$.  For each object the  previous spectral classification as given in Simbad is listed, as well as the classification adopted here. Atmospheric composition is also given, based on our model analyses. Of these 179 stars, 17 were previously identified only as ``High proper motion star'' or ``D:''  in Simbad, and we reclassify one 
from ``M7.5'' to DA (Table 4). An additional eight WDs are reclassified from DC to DA either based on new spectra (Table 4) or reexamination of existing spectra. Table 9 lists these 26 WDs which have significantly revised types.

\clearpage
%%%%%%%%%%%%%%%%%%%%%%% TABLE: Adopted Classification %%%%%%%%%%%%%%%%%%%%%%%%%%%%%%%%
\startlongtable
\begin{deluxetable}{lcccc}
\tabletypesize{\scriptsize}
\tablecaption{White Dwarf Spectral Classification
\label{tab:SpClass}}
\tablehead{
\colhead{WD Number}  & \colhead{Name} & \multicolumn{2}{c}{Spectral Type} & \colhead{Composition} \\
\colhead{}  & \colhead{}  & \colhead{Simbad}  & \colhead{Adopted}  & \colhead{} 
}
\startdata
2359$+$636      &       LSR J0002$+$6357        &       DC      &       DC10.4  &     unconstrained        \\
0002$+$729      &       GD 408  &       DB4     &       DBA3.7\tablenotemark{1} &       He      \\
0003$-$103      &       PHL 657 &       DQ      &       DQhot2.4\tablenotemark{2}       &       N(He)/N(C)$=10^{-3.0}$   \\
0003$+$177      &       G131-19 &       DQ6     &       DQ6.7   &       N(C)/N(He)$=10^{-5.62}$  \\
0015$+$004      &       PB 5847 &       DBH     &       DBP3.5\tablenotemark{3} &       He      \\
0042$-$064      &       PHL 6585        &       DC      &       DC10.1  &       N(He)/N(H)$=10^{-0.06}$  \\
0102$+$210A     &       LHS 5023        &       DC9     &       DA9.5   &       H       \\
0102$+$210B     &       LHS 5024        &       DC12    &       DC10.5  &       H       \\
\enddata 
\tablenotetext{1}{Hydrogen is a trace element in these WDs (Rolland et al. 2018).}\vskip -0.05in
\tablenotetext{2}{See Dufour (2011) for a description of the different types of DQ stars.}\vskip -0.05in
\tablenotetext{3}{These WDs are polarized, i.e. magnetic which would otherwise be indicated by the letter ``H''.}\vskip -0.05in
\tablenotetext{4}{0239$+$167: classified DQ by Kawka \& Vennes (2012), no spectrum available, will be analyzed in future paper.}\vskip -0.05in
\tablenotetext{5}{The DZ model atmospheres include trace H as well as Ca, and in all cases  N(H)/N(He)$=10^{-5.0}$.}\vskip -0.05in
\tablenotetext{6}{0714$+$458: Classified as a DQ by Weidemann \& Koester (1995), the SED is well reproduced by a pure-He atmosphere although C will be a trace element.} \vskip -0.05in
\tablenotetext{7}{Proposed oxygen-rich classification is DS, K. Williams private communication 2018.}\vskip -0.05in
\tablenotetext{8}{1704$+$481A,B: A triple system made up of a close pair of WDs and a distant  WD; the unresolved pair are both low-mass WDs (Andrews et al. 2015, Maxted et al. 2000).}\vskip -0.05in
\end{deluxetable}

\begin{deluxetable}{lccc}[b]
%\tabletypesize{\small}
%\tabletypesize{\scriptsize}
\tabletypesize{\footnotesize}
%\tablecolumns{9}
\tablewidth{0pt}
%\rotate
\tablecaption{Reclassified White Dwarfs}
\tablehead{
\colhead{WD Number}  & \colhead{Name} & \multicolumn{2}{c}{Spectral Type}  \\
\colhead{}  & \colhead{}  & \colhead{Simbad}  & \colhead{Adopted} 
}
\startdata
0102$+$210A     &       LHS 5023        &       DC9             &       DA9.5  \\
0112$-$018      &       LP 587-16       &       DC9             &       DA9.5    \\
0343$+$247      &                       &       D:              &       DC14.4  \\
0402$+$265      &       LHS 1625a       &       High PM star    &       DC12.7  \\
0500$+$573      &       LP 85-9         &       High PM star    &       D:6.0  \\
0527$+$185      &       LP 517-35       &       High PM star    &       D:9.3    \\
0530$+$054      &       LP 538-4        &       High PM star    &       D:9.7   \\
0531$+$572      &       LP 85-33        &       High PM star    &       DC:10.2  \\
\enddata 
\end{deluxetable}

\begin{figure}
\vskip -1.8in
\begin{center}
    \includegraphics[angle=0,width=0.9\textwidth]{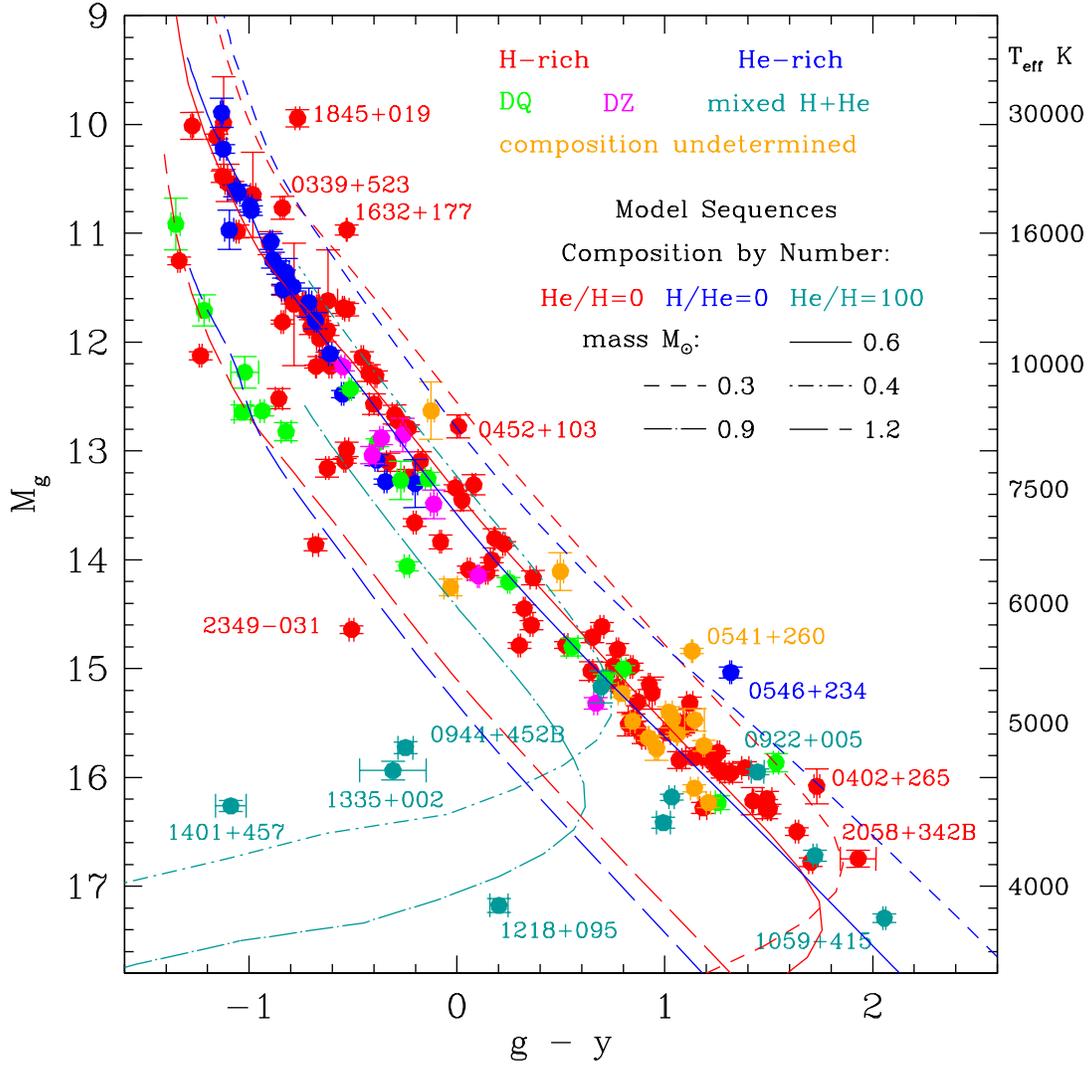}\\
\vskip -0.5in
\caption{Symbols represent the core sample of 179 WDs with NOFS parallaxes. Lines are iso-mass model sequences. The five line types indicate mass,  and the line color indicates atmospheric composition, as shown in the legend.
%Symbol colors indicate atmospheric composition as determined by model fits. Sequences are pure-H and pure-He atmospheres with surface gravities as indicated in the legend. 
Values of $T_{\rm eff}$ for the mass$=$0.6 $M_{\odot}$ models are indicated on the right axis. Outliers and candidate binary systems  (Section 8.2) are identified.
}
\end{center}
\end{figure}

\begin{figure}[t]
\vskip -0.1in
\begin{center}
    \includegraphics[angle=-90,width=1.0\textwidth]{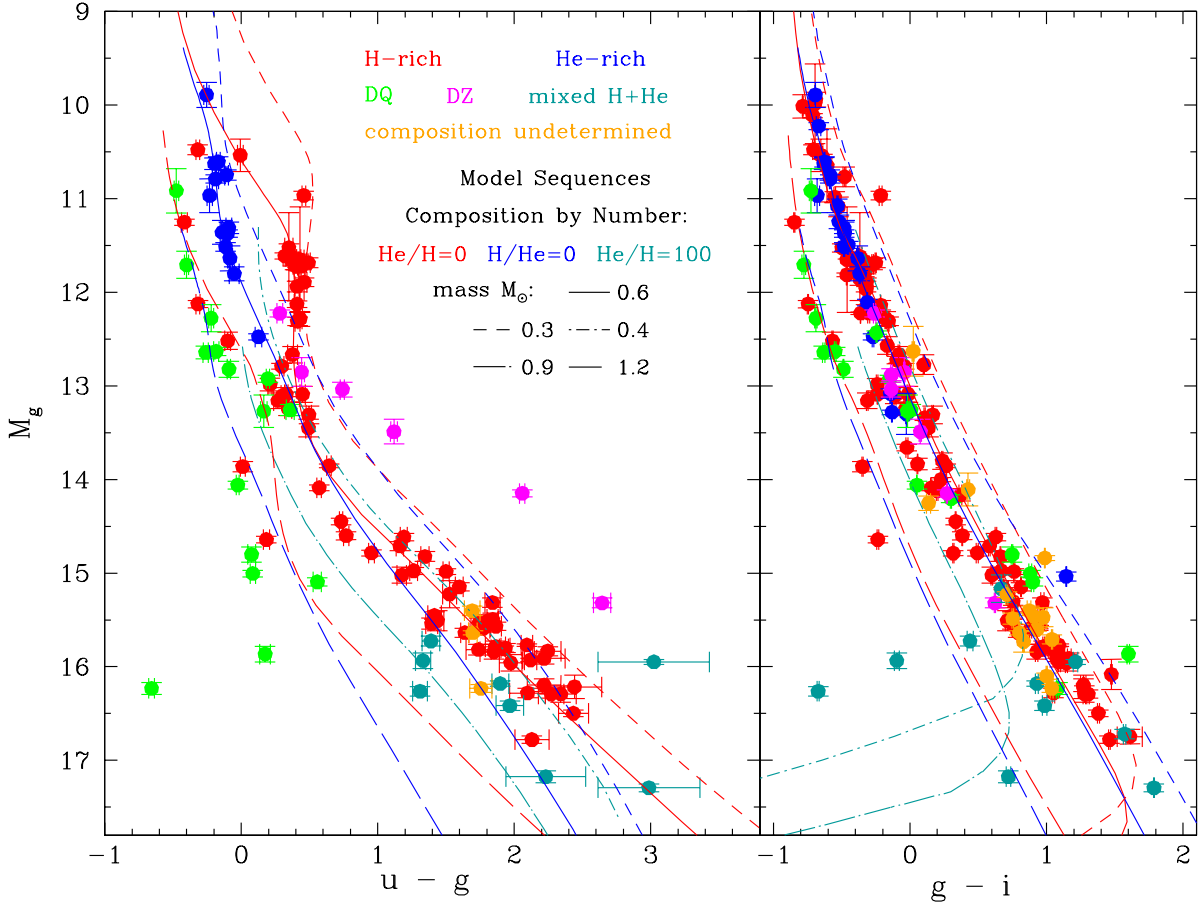}\\
\vskip -0.2in
 \caption{Optical  color-magnitude diagrams.   Symbol colors and line types are as given in the legend.
}
\end{center}
\end{figure}

Figure 8 shows the 179 stars in the $g - y:M_g$ color-magnitude diagram, with outliers identified. Figure 9 shows $u - g:M_g$ and $g - i:M_g$ and Figure 10 shows  $J - K:M_J$ and  $J -$ W2 or [4.5] against $M_J$. Some stars have both W2 and [4.5] measurements and are plotted twice in Figure 10. Model sequences are also shown in the Figures. In Figure 10 the model sequence is for W2 only; the models show a difference of $\lesssim 5$\% between  W2 and [4.5] for our sample.

Stars that appear over-luminous are either low-gravity and low-mass WDs, binaries (e.g. Figure 6), or are unusually red.  In Section 8.2 we discuss the mass distribution of the sample and identify new candidate unresolved binary systems. Stars that appear sub-luminous are either high-gravity and high-mass WDs or are unusually blue. 
One hydrogen-rich WD that stands out in Figure 8  is 2349$-$031; 
this WD is the most massive in our sample with a mass $1.326 \pm 0.012~M_{\odot}$ (Section 8).

Figures 8 and 9 show that the cool mixed hydrogen plus helium atmosphere WDs can be extremely blue in $g - y$ and $g - i$. These objects
generally show strong pressure-induced H$_2$ absorption in the far-red and near-infrared, leading to blue $g - y$, $g - i$  and $J - K$, as well as faint $M_J$ values \citep[Figures 8 --10]{Jorgensen2000}. Figure 10 demonstrates that $J - K$ diverges for hydrogen plus helium atmospheres cooler than $T_{\rm eff} \approx 4500$~K,  due to the strong pressure-induced H$_2$ absorption at $\lambda \sim 2~\mu$m in cool pure-hydrogen atmospheres \citep{Bergeron1995}.

\begin{figure}[!t]
\vskip -0.1in
\begin{center}
    \includegraphics[angle=-90,width=1.0\textwidth]{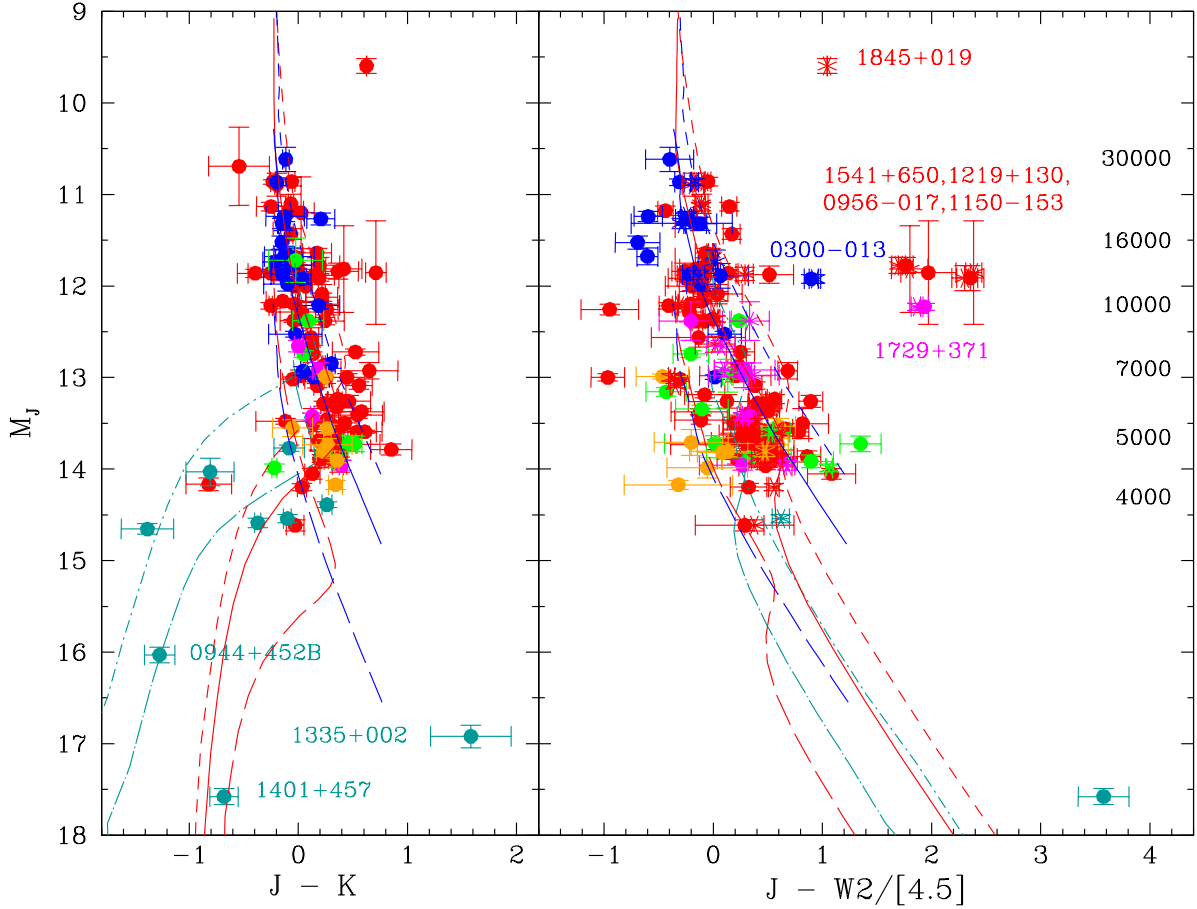}\\
\vskip -0.3in
 \caption{Infrared and mid-infrared color-magnitude diagrams for our core sample.   Symbol colors and line types are as in Figures 8 and 9.  $T_{\rm eff}$ for mass $=$ 0.6 $M_{\odot}$  are  indicated  on the right axis.
}
\end{center}
\end{figure}

\begin{figure}[!t]
\vskip -0.1in
\begin{center}
    \includegraphics[angle=0,width=0.65\textwidth]{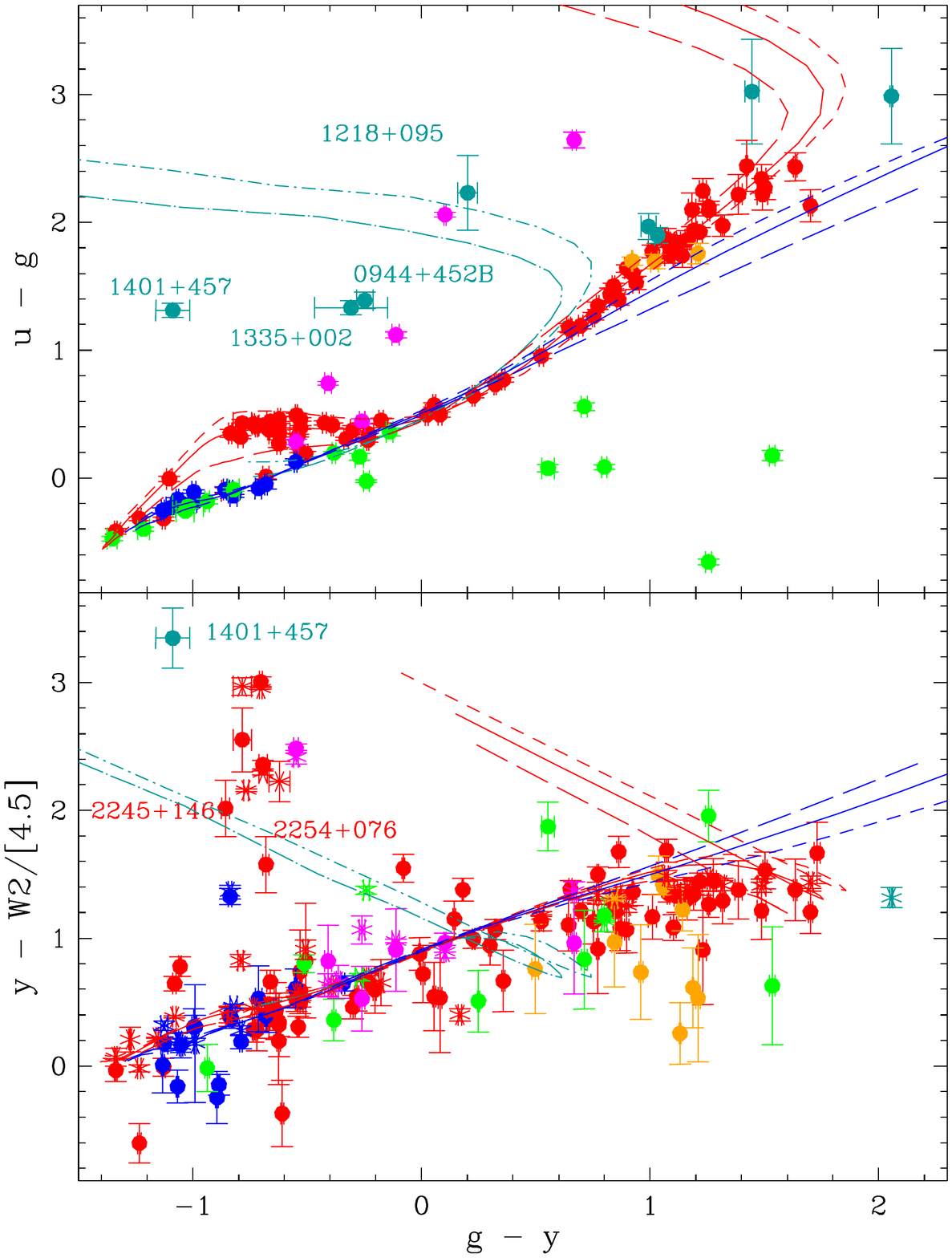}\\
\vskip -0.5in
\caption{Blue to red, and green to mid-infrared, color-color plots for our core sample. Symbol colors and line types are as in Figures 8 and 9. 
%In the lower panel two stars are identified which have an apparent mid-infrared excess, the seven stars identified in the right panel of Figure 10 also show an excess in this plot.
}
\end{center}
\end{figure}

Figure 9 shows that the $u - g$ color is a particularly useful indicator of atmospheric composition.
DQ stars have carbon absorption bands at $\lambda \sim 450$~nm  \citep[Figure 7]{Wesemael1993,Gentile2018}, and hence have blue $u - g$ colors. DZ stars generally show calcium absorption at $\lambda \sim 390$~nm \citep{Wesemael1993} and so are red in $u - g$. The $u - g$ color also separates DA from DB stars for $ 10000 \lesssim T_{\rm eff}$~K  $ \lesssim 16000$, where DAs have 
strong hydrogen absorption in the $u$-band.

Figure 11 shows blue to red and green to mid-infrared color-color plots. The $g - y:u - g$ plot shows some of the features already highlighted:  the significant separation in $u - g$ for DAs and DBs with  $ 10000 \lesssim T_{\rm eff}$~K  $ \lesssim 16000$, the bluer and redder $u - g$ colors of DQ and DZ stars respectively, and the blue $g - y$ colors of cool mixed H and He WDs. The $g - y:y -$ W2/[4.5] plot in Figure 11 and the $J -$ W2/[4.5]:$M_J$ plot in Figure 10 shows that there are WDs in the sample with mid-infrared flux excesses; we discuss these further in the next sub-section. 

White dwarfs cooler than $\sim$5000~K are generally featureless, apart from the broad pressure-induced H$_2$ features in hydrogen-rich atmospheres. In some cases the SED allows us to determine the composition of the atmospheres (Figure 3) but in others the data coverage or precision, or the particular combination of parameters, leaves the composition unconstrained. 
\citet{Kowalski2006} compared their models to observed color-color diagrams and concluded that most cool DC stars are hydrogen-rich. The trends we see in Figures 9 and 11 support that conclusion.

\begin{deluxetable}{llllc}[!b]
\tablecolumns{5} 
%\movetabledown=1.25in
%\def\arraystretch{0.8} 
%\tabletypesize{\small}
\tabletypesize{\scriptsize}
%\tablecolumns{9}
\tablewidth{0pt}
%\rotate
\tablecaption{White Dwarfs with Mid-Infrared Excess Flux}
\tablehead{
\colhead{WD Number}   & \colhead{Name} & \colhead{Sp. Type} & \colhead{Reference} & \colhead{Notes} 
}
\startdata
%\hline
%\cutinhead{Known Disks}
0300$-$013   & GD 40 &  DBZ3.6 &  \citet{Jura2007} & Known disk\\
0956$-$017  & EQ  J0959$-$0200 & DAZ4.0 &    \citet{Girven2011} & Known disk\\
1150$-$153  &   & DA4.0 &   \citet{Kilic2007} & Known disk\\
1219$+$130    & SDSS  J1221$+$1245  & DAZ4.2 &   \citet{Girven2011} &  Known disk\\
1541$+$650  &   V*  KX Dra & DA4.5   & \citet{Kilic2012} & Known disk \\
1729$+$371  &  GD 362 & DAZB4.9  & \citet{Becklin2005,Kilic2005} & Known disk\\
%\cutinhead{Previously Suspected  Red Companions}
1845$+$019  &         & DA2.4 &     candidate DA$+$dM \citet{Hoard2007} & SED supports red dwarf companion\\
2245$+$146   &    & DAP3.2   & this work & SED suggests circumstellar disk   \\
2254$+$076  &  G28-27 & DAH4.2 &   candidate DA$+$M0 \citet{Debes2011} & SED supports disk, not companion\\
%\cutinhead{Newly Identified}
\enddata 
\end{deluxetable}

\vskip 1.0in
\subsection{White Dwarfs with Mid-Infrared Excess Flux}

Seven WDs are identified in the $M_J:J -$ W2/[4.5] plot (Figure 10) that are very red in $J -$ W2/[4.5]. An additional two WDs are identified in Figure 11 that are red in $y -$ W2/[4.5]. Table 10 lists these nine WDs which have significant mid-infrared excess. Six of them are WDs with previously known circumstellar dust disks. Two have been flagged in the literature as possibly having red companions and the last has not previously been recognized to have an infrared excess to our knowledge. We compared the SEDs of these three WDs to those with similar type known to have red companions or disks in Figure 12. The WDs with red dwarf companions show excess flux in the near-infrared as well as the mid-infrared, while WDs with lower temperature dust disks have a mid-infrared flux excess only.
We find that the SED of 1845$+$019  matches the SED of a WD with an M or L dwarf companion, as previously suggested by \citet{Hoard2007} (note that 1845$+$019 appears red in $g - y$ in Figure 8). However the SED of 2254$+$076 does not look like a WD with a red companion, as previously suggested by \citet{Debes2011}, instead it looks like a WD with a dust disk.  2245$+$146 also appears to have a dust disk although near-infrared photometry would help to secure this.

\begin{figure}[!b]
%\begin{center}
\hskip -1.8in
\vskip -0.8in
    \includegraphics[angle=0,width=0.57\textwidth]{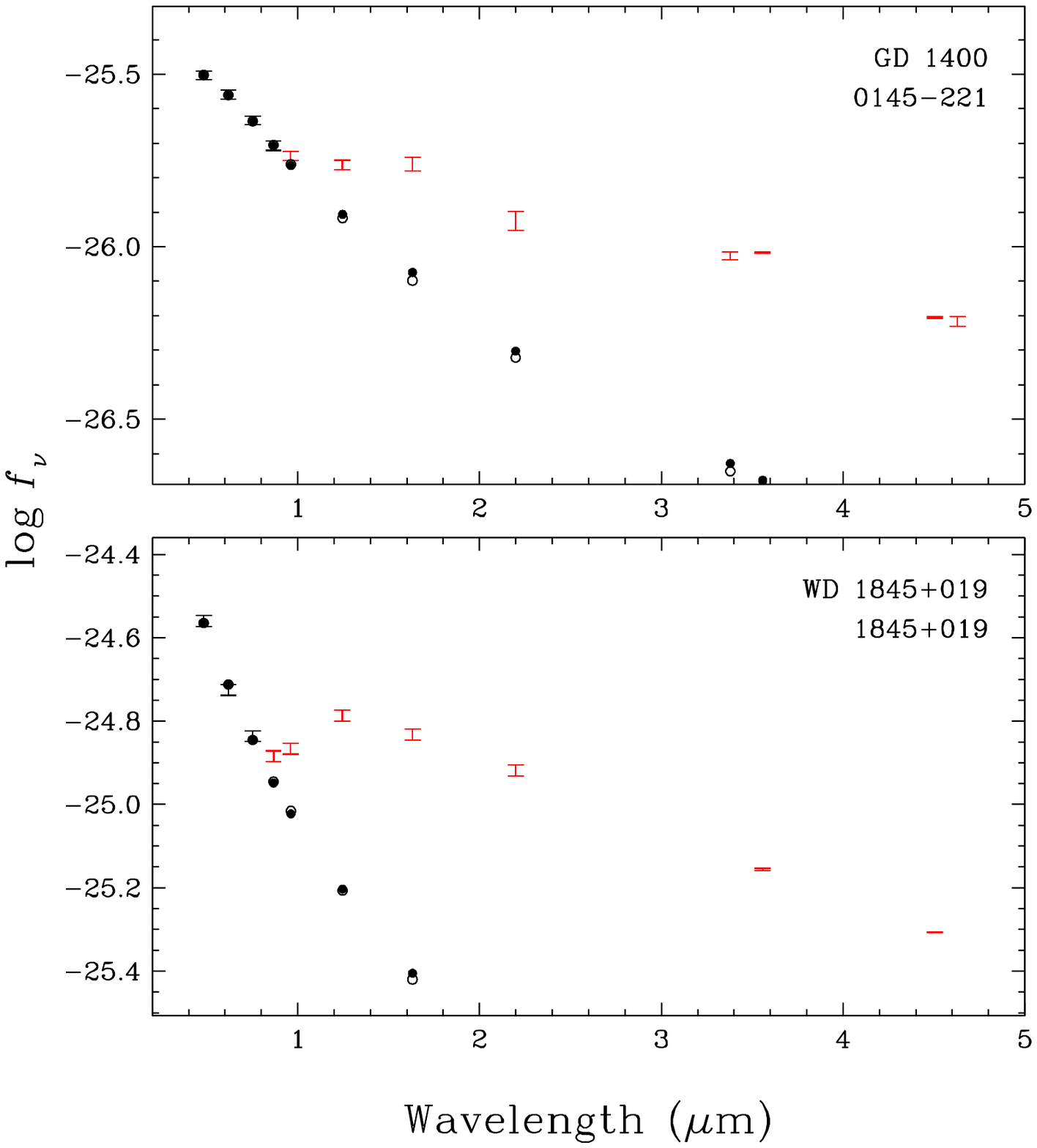}\\
\vskip -4.48in
\hskip 3.3in
    \includegraphics[angle=0,width=0.57\textwidth]{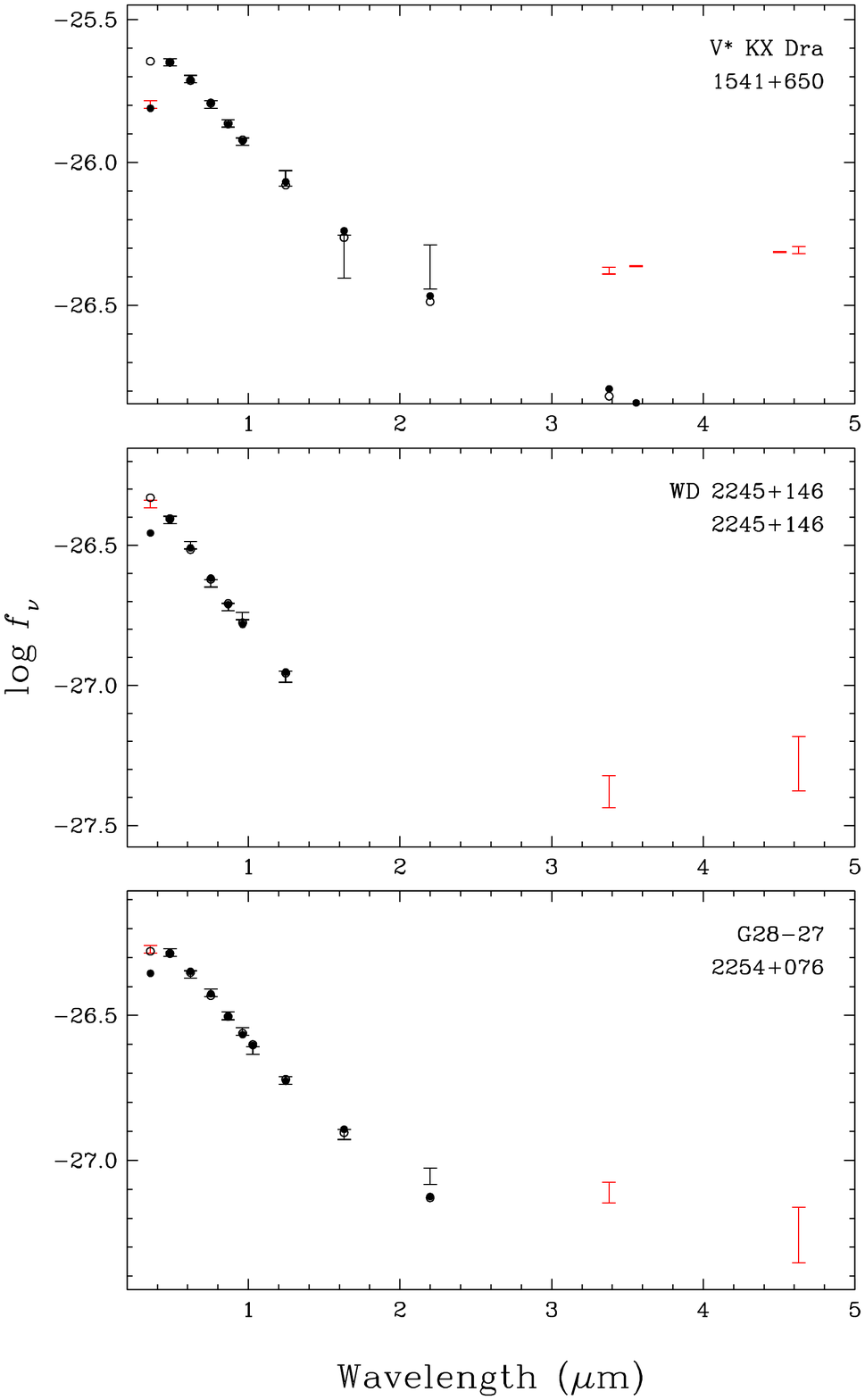}
%\vskip -1.7in
%    \includegraphics[angle=0,width=0.75\textwidth]{Disks.pdf}\\
\vskip -0.2in
\caption{The  left panel compares the SED of 0145$-$221, a DA4.6 with an L dwarf companion (Table 6), to that of 1845$+$019, confirming that the latter has a red dwarf companion.  The right panel compares the SED of 1541$+$650, a DA4.5 with a dust disk (Table 10) to those of  2245$+$146 and 2254$+$076, which we suggest also have disks.
}
%\end{center}
\end{figure}

\clearpage
\section{Physical Properties of the Sample of 179  White Dwarfs}

\subsection{Properties by Atmospheric Composition}

Tables 11 -- 15 list the derived parameters and their uncertainties for each WD, grouped by atmospheric compositions: pure hydrogen, pure helium, the metal-rich DQ and DZ, the cool  mixed hydrogen-helium and the unconstrained atmospheres. The $T_{\rm eff}$, log $g$, mass and cooling age are determined by the model fits as described in Section 5. For the WDs with unknown composition we adopted the average of the pure-hydrogen and pure-helium parameters, and included the range of solutions in the uncertainties. Uncertainties in the photometry are propagated through the model derived parameters. 

Our sample of DA and DB WDs overlaps with those of \citet{Gianninas2011} and \citet{Rolland2018}. Those authors used the same models used in this study to determine WD atmospheric properties spectroscopically, while we used the photometric technique together with updated parallaxes. 
Most of our WDs have pure-hydrogen atmospheres, and so we compared the properties determined for 49 DA stars in common with \citet{Gianninas2011}. The $T_{\rm eff}$ values agree for WDs with $T_{\rm eff} < 10000$~K, but for the warmer stars the spectroscopic temperatures are higher by $\sim 700$~K.  
Similarly, the  gravity and mass determinations agree within $2~\sigma$ ($\delta$(log~$g)\approx~0.1$ and $\delta$(mass)$\approx~0.05~M_{\odot}$) for 60\% of the sample but for 35\% of the sample the spectroscopic values  are higher,  with $\delta$(log~$g)\approx~0.3$ and $\delta$(mass)$\approx~0.2~M_{\odot}$. The high values found for these parameters by the spectroscopic method has been noted previously  
\citep[e.g.][]{Bergeron1990a, Eisenstein2006, Genest2014, Kepler2007, Limoges2015, Tremblay2015}
and has  been attributed to shortcomings in the modelled line broadening. Recently, three-dimensional pure-hydrogen model atmospheres have been calculated \citep{Tremblay2013} which provide a correction to the one-dimensional model-derived parameters. Figure 4 of Tremblay et al.  
shows that for DAs with $T_{\rm eff} > 10000$~K the correction to $T_{\rm eff}$ is $\sim -500$~K, and for DAs with $7000 \lesssim T_{\rm eff}$~K~$\lesssim 12000$  the correction to log $g$ is $\sim -0.2$~dex.
Application of these corrections bring the  $T_{\rm eff}$ values for all but one WD within  $2~\sigma$, and excluding four very discrepant stars, all log $g$ values then agree within $2.5~\sigma$. One of the log $g$ discrepant stars is a magnetic WD where the spectroscopic fit is poor  \citep[1658$+$440,][]{Gianninas2011}. Two of the three other very log $g$ discrepant WDs are identified as candidate binary systems in Section 8.3 (0452$+$103, 1632$+$177) and it is possible that the third object, 1310$+$583, is an unresolved binary also. 
Hence the comparison to the spectroscopic analysis of DAs by \citet{Gianninas2011} indicates that our results are valid.

The total age of each WD was estimated for the WDs more massive than 0.5~$M_{\odot}$.
Less massive WDs cannot have formed as a normal single WD \citep{Bergeron1992}. These systems may be unresolved multiple systems in which case an analysis assuming a single WD will produce too large a radius and too small a mass. The systems may also be very close binaries which have evolved through a common-envelope stage which can produce low mass WDs 
\citep{Iben1993}.

The initial mass $M_i$ was estimated from the WD final mass $M_f$ using the initial-final mass relations from  
\citet{Kalirai2009} and \citet{Cummings2016} for WDs with masses  0.5 -- 0.7~$M_{\odot}$ and $>$ 0.7  ~$M_{\odot}$, respectively: 
\vskip -0.15in
$$ M_f = (0.109 \pm 0.007)\times M_i + (0.428 \pm 0.025)$$ 
\vskip -0.05in
\noindent
for  $0.5 \leq M_f \leq 0.7$ and
\vskip -0.15in
$$ M_f = (0.143 \pm 0.005)\times M_i + (0.294 \pm 0.020)$$ 
\vskip -0.05in
\noindent
for  $M_f > 0.7$. The time spent on the main sequence as a function of $M_i$ was determined from the \href{http://people.sissa.it/~sbressan/parsec.html}{PARSEC stellar  evolutionary code} \citep{Bressan2012}. Using solar metallicity tracks, a fit to the main sequence lifetime $MS_t$ as a function of initial stellar mass $M_i$ gives 
\vskip -0.1in
$$ MS_t = (9.377 \pm 0.022)10^9 \times M_i^{(-2.715 \pm 0.042)}$$ 
\vskip -0.05in
This time was added to the cooling age to give the total age in Tables 11 -- 15. Uncertainties were propagated through the mass and age calculations.

% initial mass from Cummings et al. 2016 ApJ 818 for final mass >0.7 
% mfinal=(0.143 +- 0.005)Minit + 0.294 +- 0.020
% and Kalirai et al. 2009 ApJ for mass 0.5-0.7
% mfinal=(0.109 +- 0.007)Minit + 0.428 +- 0.025
% MS age from Girardi et al 2000 isochrones solar metallicity Z=0.019, Y=0.273

\subsection{Mass}

Figure 13 shows that the sample has the typical peak in mass around 0.6~$M_{\odot}$ 
\citep{Giammichele2012,Limoges2015}.
\citet{Kilic2018} uses a large volume-limited sample of WDs with {\it Gaia} parallaxes to show that there are a significant number of WDs with mass $\sim 0.8 ~M_{\odot}$ which likely formed through mergers. Our smaller and more inhomogeneous sample does not show evidence of this bump in the mass distribution.

In our sample of 179 notionally single WDs, twelve have masses below 0.45~$M_{\odot}$ (allowing for the uncertainty in mass). These  are possibly unresolved binary systems, and are listed in Table 16 for future followup such as radial velocity or variability studies. The twelve WDs are also identified in the color-magnitude diagram of Figure 8.

\begin{figure}[!b]
\vskip -0.1in
\begin{center}
    \includegraphics[angle=-90,width=0.85\textwidth]{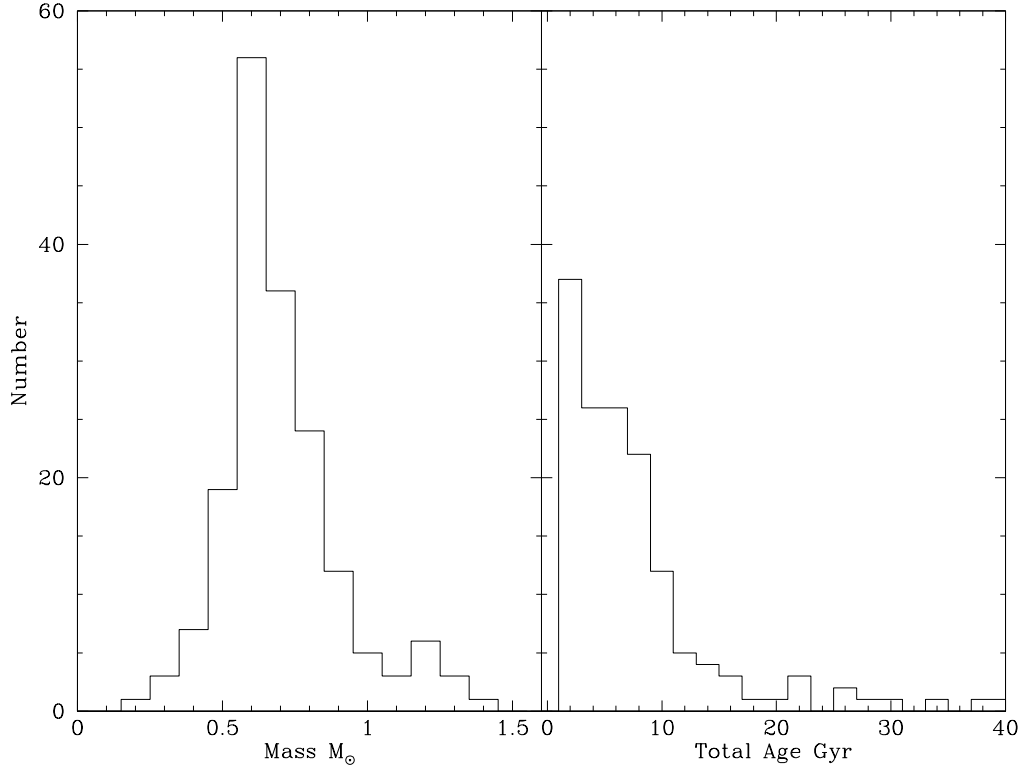}\\
\vskip -0.2in
\caption{Histograms of WD mass as determined by our model fits (left) and total age (right) which is the sum of the WD cooling age and the progenitor's time on the main sequence. The typical uncertainty in mass is 0.04 $M_{\odot}$. The uncertainty in age can be as large as 1000~Gyr, but the typical uncertainty is 2~Gyr.
}
\end{center}
\end{figure}

\begin{longrotatetable}
\begin{deluxetable}{lcr@{ $\pm$ }lr@{ $\pm$ }lr@{ $\pm$ }lr@{ $\pm$ }lr@{ $+$ }l@{ $-$ }lr@{ $+$ }l@{ $-$ }lr}
%\tabletypesize{\tiny}
\tabletypesize{\scriptsize}
\tablecaption{Properties of WDs with Hydrogen-Rich Atmospheres
\label{tab:Hprops}}
\tablehead{
\colhead{WD Number}  &  \colhead{Spectral} &  \multicolumn{2}{c}{Parallax} & \multicolumn{2}{c}{$T_{\rm eff}$} & \multicolumn{2}{c}{Mass} 
& \multicolumn{2}{c}{log $g$} & \multicolumn{3}{c}{Cooling Age} &  \multicolumn{3}{c}{Total Age} & \colhead{$V_{\rm tan}$}  \\
\colhead{}  & \colhead{Type}  & \multicolumn{2}{c}{mas} & \multicolumn{2}{c}{K}  & \multicolumn{2}{c}{$M_{\odot}$} & \multicolumn{2}{c}{dex} &
 \multicolumn{6}{c}{Myr} & \colhead{km~s$^{\rm -1}$} 
}
\startdata
0102$+$210A & DA9.5 & 29.31 & 0.79 & 5278 &  89 & 0.777 & 0.039 & 8.301 & 0.042 & 7101 & 461 & 494 & 7445 & 345 & 300 & 78.5\\
0102$+$210B & DC10.5 & 29.31 & 0.79 & 4818 & 78 & 0.620 & 0.041 & 8.067 & 0.046 & 7158 & 488 & 533 & 9174 & 4783 & 781 &  79.8\\
0112$-$018 & DA9.5 & 8.60 & 0.53 & 5311 & 73 & 0.752 & 0.028 & 8.265 & 0.030 & 6712 & 384 & 418 & 7110 & 262 & 230 &  91.7\\
0114$-$049A & DC:10.3 & 24.55 & 0.55 & 4885 & 75 & 0.550 & 0.036 & 7.951 & 0.043 & 5650 & 594 & 609 & 12556 & 43857 & 4394  & 73.0\\
0145$-$174 & DA6.8 & 13.12 & 0.56 & 7403 & 169 & 0.552 & 0.048 & 7.929 & 0.059 & 1192 & 128 & 119 & 7799 & 76090 & 4909  & 400.9\\
0202$-$055 & DC12.2 & 34.12 & 0.60 & 4124 & 52 & 0.596 & 0.032 & 8.031 & 0.036 & 8956 & 305 & 333 & 11853 & 7084 & 1498 &  144.1\\
0203$+$183 & DC11.6 & 27.56 & 0.84 & 4329 & 83 & 0.493 & 0.046 & 7.857 & 0.057 & 6640 & 678 & 687  &  44801    & \nodata &   33273 & 135.7\\
0228$+$269 & DC:10.4 & 35.63 & 1.08 & 4855 & 77 & 0.513 & 0.042 & 7.888 & 0.052 & 5068 & 683 & 681 & 23489 & \nodata &  14619 & 56.8\\
\enddata 
\tablenotetext{a}{Relative parallax, absolute parallax will be larger.}
\end{deluxetable}
\end{longrotatetable}

\begin{longrotatetable}
\begin{deluxetable}{lcr@{ $\pm$ }lr@{ $\pm$ }lr@{ $\pm$ }lr@{ $\pm$ }lr@{ $+$ }l@{ $-$ }lr@{ $+$ }l@{ $-$ }lr}
%\tabletypesize{\tiny}
\tabletypesize{\scriptsize}
\tablecaption{Properties of WDs with Helium-Rich Atmospheres
\label{tab:Heprops}}
\tablehead{
\colhead{WD Number}  &  \colhead{Spectral} &  \multicolumn{2}{c}{Parallax} & \multicolumn{2}{c}{$T_{\rm eff}$} & \multicolumn{2}{c}{Mass} 
& \multicolumn{2}{c}{log $g$} & \multicolumn{3}{c}{Cooling Age} &  \multicolumn{3}{c}{Total Age} & \colhead{$V_{\rm tan}$}  \\
\colhead{}  & \colhead{Type}  & \multicolumn{2}{c}{mas} & \multicolumn{2}{c}{K}  & \multicolumn{2}{c}{$M_{\odot}$} & \multicolumn{2}{c}{dex} &
 \multicolumn{6}{c}{Myr} & \colhead{km~s$^{\rm -1}$} 
}
\startdata
0002$+$729 & DBA3.7 & 28.35 & 0.49 & 13537 & 733 & 0.610 & 0.046 & 8.040 & 0.055 & 290 & 48 & 58 & 2621 & 7451 & 1503 & 38.0 \\
0015$+$004 & DBP3.5 & 7.55 & 0.44 & 14532 & 580 & 0.621 & 0.063 &  8.054 & 0.075 & 240 & 45 & 39 & 2228 & 9950 & 1344 & 17.1 \\
0112$+$104 & DB1.8 & 8.60 & 0.53 & 27861 & 1507 &       0.517 & 0.047 &         7.816 & 0.072 & 13 & 2 & 3 & 16271 & \nodata & 13630 & 27.3 \\
0114$-$049B & DC:9.8 & 24.55 & 0.55 & 4979 &  54 &  0.535 &  0.029 &  7.945  & 0.035  & 5250  &  293  &  324  & 15111 &  64567 & 6897 & 76.1\\
0300$-$013 & DBZ3.6 & 16.61 & 0.35 & 13921 & 428 & 0.651 & 0.032 & 8.106 & 0.037 & 299 & 31 & 34 & 1641 & 2057 & 710 & 27.6 \\
0541$+$260 & DC10.0 & 28.21 & 0.32 & 5050  & 51 &  0.306 &  0.017 & 7.475  & 0.030 & 2225 &  85 & 79 & \nodata & \nodata & \nodata & 286.4 \\
0546$+$234 & D:10.6 & 17.33 & 0.39 & 4746 & 40 & 0.236 & 0.018 & 7.274 & 0.041 & 2160 & 79 & 81 & \nodata & \nodata & \nodata & 375.6 \\
0637$+$335 & DC5.7 & 15.76 & 1.59 & 8336 & 276 &  0.739 & 0.103 &  8.255 & 0.114  & 1501 & 442 & 328  & 1931 & 2174  & \nodata  & 55.5 \\
\enddata 
\end{deluxetable}
\end{longrotatetable}

\begin{longrotatetable}
\begin{deluxetable}{lcr@{ $\pm$ }lr@{ $\pm$ }lr@{ $\pm$ }lr@{ $\pm$ }lr@{ $+$ }l@{ $-$ }lr@{ $+$ }l@{ $-$ }lr}
%\tabletypesize{\tiny}
\tabletypesize{\scriptsize}
\tablecaption{Properties of WDs with Metal-Rich Atmospheres
\label{tab:QZprops}}
\tablehead{
\colhead{WD Number}  &  \colhead{Spectral} &  \multicolumn{2}{c}{Parallax} & \multicolumn{2}{c}{$T_{\rm eff}$} & \multicolumn{2}{c}{Mass} 
& \multicolumn{2}{c}{log $g$} & \multicolumn{3}{c}{Cooling Age} &  \multicolumn{3}{c}{Total Age} & \colhead{$V_{\rm tan}$}  \\
\colhead{}  & \colhead{Type}  & \multicolumn{2}{c}{mas} & \multicolumn{2}{c}{K}  & \multicolumn{2}{c}{$M_{\odot}$} & \multicolumn{2}{c}{dex} &
 \multicolumn{6}{c}{Myr} & \colhead{km~s$^{\rm -1}$} 
}
\startdata
0003$-$103 & DQhot2.4 & 6.20 & 0.41 & 20882 & 1103 & 1.125 & 0.046 & 8.863 & 0.075 & 333 & 72 & 53 & 412 & 52 & 23 & 63.4\\
0003$+$177 & DQ6.7 & 18.40 & 1.16 & 7468  &  66  & 0.535 & 0.029 &  7.932 &  0.035 & 1213 & 60 & 59 & 11074  & 64832 &  7085 & 115.5\\
0203$+$207 & DQA3.1 & 11.93 & 0.38 & 16150 &  612  & 1.205 & 0.021 &  9.038 &  0.044 &  966 &  91 &  88  & 1027 & 79 & 73  & 130.3\\  
0239$+$167\tablenotemark{a} & DQ7.7 & 25.21 & 0.47 & \nodata & \nodata & \nodata & \nodata & \nodata & \nodata & \nodata & \nodata & \nodata & \nodata & \nodata & \nodata & 133.2\\
0246$+$734 & DZ6.5 & 13.14 & 0.93 & 7709 & 106 & 0.444 & 0.063 & 7.762 & 0.088 & 902 & 105 & 93 & \nodata & \nodata & \nodata & 86.0\\
0625$+$100 & DZ5.9 & 18.36 & 0.54 & 8519 & 96 & 0.603 & 0.031 & 8.042 &  0.036 &  1002  & 57 & 55 & 3595  &  6101 & 1505 & 37.4\\
0714$+$458 & DQ5.0 & 28.30 & 0.22 & 10007  & 393  &  0.652 & 0.040 & 8.117 &  0.045 & 742 & 98 & 86 &  2068 & 2438 & 679 & 35.0\\
0913$+$103 & DQ5.9 & 26.72 & 0.45 & 8534 & 62 & 0.612 & 0.017 & 8.058 & 0.020 & 1018 & 34 & 33 & 3281 & 3119 & 1148 & 33.7\\
\enddata 
\tablenotetext{a}{Analysis postponed to future work}
\end{deluxetable}
\end{longrotatetable}

\begin{longrotatetable}
\begin{deluxetable}{lcr@{ $\pm$ }lr@{ $\pm$ }lr@{ $\pm$ }lr@{ $\pm$ }lr@{ $+$ }l@{ $-$ }lr@{ $+$ }l@{ $-$ }lr}
%\tabletypesize{\tiny}
\tabletypesize{\scriptsize}
\tablecaption{Properties of Cool WDs with Mixed Hydrogen and Helium Atmospheres
\label{tab:mixprops}}
\tablehead{
\colhead{WD Number}  &  \colhead{Spectral} &  \multicolumn{2}{c}{Parallax} & \multicolumn{2}{c}{$T_{\rm eff}$} & \multicolumn{2}{c}{Mass} 
& \multicolumn{2}{c}{log $g$} & \multicolumn{3}{c}{Cooling Age} &  \multicolumn{3}{c}{Total Age} & \colhead{$V_{\rm tan}$}  \\
\colhead{}  & \colhead{Type}  & \multicolumn{2}{c}{mas} & \multicolumn{2}{c}{K}  & \multicolumn{2}{c}{$M_{\odot}$} & \multicolumn{2}{c}{dex} &
 \multicolumn{6}{c}{Myr} & \colhead{km~s$^{\rm -1}$} 
}
\startdata
0042$-$064 & DC10.1 & 33.04 & 0.40 & 4983 & 80 & 0.890 & 0.029 & 8.486 & 0.032 & 7120 & 106 & 100 & 7315 & 55 & 25 &  98.2\\
0343$+$247 & DC14.4 & 25.75 & 0.59 & 3506 & 39 & 0.320 & 0.027 & 7.519 & 0.046 & 5418 & 419 & 387 & \nodata & \nodata & \nodata  & 233.2\\
0922$+$005 & D:12.2 & 15.67 & 0.16 & 4127 & 65 & 0.338 & 0.024 & 7.559 & 0.040 & 4018 & 343 & 303 & \nodata & \nodata & \nodata &  175.6\\
0944$+$452B & DC16.7 & 18.54 & 0.42 & 3009 & 27 & 0.125 & 0.024 & 6.864 & 0.089 & 3072 & 308 & 254 & \nodata & \nodata & \nodata  & 22.5\\
1059$+$415 & DC14.9 & 28.85 & 0.54 & 3371 & 36 & 0.354 & 0.025 & 7.598 & 0.040 & 6640 & 355 & 360 & \nodata & \nodata & \nodata  & 286.9\\
1218$+$095 & DC14.3 & 24.94 & 0.72 & 3531 & 37 & 0.893 & 0.033 & 8.492 & 0.037 & 8535 & 166 & 140 & 8727 & 113 & 61  & 96.4\\
1335$+$002 & DC18.2 & 19.68 & 0.76 & 2772 & 32 & 0.311 & 0.021 & 7.301 & 0.051 & 1866 & 108 & 83 & \nodata & \nodata & \nodata  & 41.5\\
1401$-$149 & DC11.1 & 20.19 & 1.37 & 4554 & 90 & 0.404 & 0.069 & 7.701 & 0.100 & 4022 & 849 & 695 & \nodata & \nodata & \nodata  & 91.4\\
\enddata 
\end{deluxetable}
\end{longrotatetable}

\begin{longrotatetable}
\begin{deluxetable}{lcr@{ $\pm$ }lr@{ $\pm$ }lr@{ $\pm$ }lr@{ $\pm$ }lr@{ $+$ }l@{ $-$ }lr@{ $+$ }l@{ $-$ }lr}
%\tabletypesize{\tiny}
\tabletypesize{\scriptsize}
\tablecaption{Properties of WDs with Unconstrained Composition Atmospheres
\label{tab:Uprops}}
\tablehead{
\colhead{WD Number}  &  \colhead{Spectral} &  \multicolumn{2}{c}{Parallax} & \multicolumn{2}{c}{$T_{\rm eff}$} & \multicolumn{2}{c}{Mass} 
& \multicolumn{2}{c}{log $g$} & \multicolumn{3}{c}{Cooling Age} &  \multicolumn{3}{c}{Total Age} & \colhead{$V_{\rm tan}$}  \\
\colhead{}  & \colhead{Type}  & \multicolumn{2}{c}{mas} & \multicolumn{2}{c}{K}  & \multicolumn{2}{c}{$M_{\odot}$} & \multicolumn{2}{c}{dex} &
 \multicolumn{6}{c}{Myr} & \colhead{km~s$^{\rm -1}$} 
}
\startdata
0306$+$663 & DC10.7 & 26.48 & 0.59 & 4707 & 78 & 0.506 & 0.041 & 7.886 & 0.051 & 5503 & 570 & 599 & 30649 & \nodata & 20907 & 145.1\\
0357$+$513 & DC10.0 & 39.26 & 0.42 & 5139 & 65 & 0.741 & 0.023 & 8.256 & 0.026 & 7023 & 251 & 269 & 7634 & 274 & 39 & 106.2 \\
0409$+$237 & DC:6.7 & 13.06 & 0.45 & 7378 & 229 & 0.935 & 0.045 & 8.542 & 0.053 & 3474 & 270 & 267 & 3637 & 220 & 189 & 160.9 \\
0500$+$573 & D:6.2 & 9.25 & 1.12 & 8226 & 262 &  0.426 & 0.101 & 7.700 & 0.154 & 706 & 138 & 106 & \nodata & \nodata & \nodata & 148.7 \\
0531$+$572 & DC:10.2 & 24.41 & 0.92 & 4903 & 72 & 0.523 & 0.047 & 7.916 & 0.057 & 5166 & 651 & 694 & 24146 & \nodata & 15587 & 53.8 \\
0541$+$750 & D:8.4 & 14.22 & 1.14 & 5892 & 140 & 0.440 & 0.078 & 7.747 & 0.110 & 1765 & 287 & 257 & \nodata & \nodata & \nodata & 85.4\\
0701$+$517 & D:9.8 & 26.86 & 1.35 & 5064 & 79 & 0.742 & 0.057 & 8.264 & 0.061 & 7315 & 326 & 424 & 7748 & 620 & 145 & 83.2 \\
0708$+$462 & DC:10.5 & 13.82 & 0.63 & 4803 & 76 & 0.463 & 0.054 & 7.806 & 0.070 & 4348 & 779 & 714 & \nodata & \nodata & \nodata & 184.7 \\
\enddata 
\end{deluxetable}
\end{longrotatetable}

\begin{deluxetable}{llcr}[!t]
\tablecolumns{5} 
%\movetabledown=1.25in
%\def\arraystretch{0.8} 
%\tabletypesize{\small}
\tabletypesize{\small}
%\tablecolumns{9}
\tablewidth{0pt}
%\rotate
\tablecaption{Candidate Unresolved White Dwarf Binaries}
\tablehead{
\colhead{WD Number}   & \colhead{Name} & \colhead{Spectral Type} & \colhead{Mass $M_{\odot}$\tablenotemark{a}}  
}
\startdata
0339$+$523 & EGGR 567 & DA3.8   & 0.37 $\pm$ 0.03 \\
0343$+$247 &          &  DC14.4  & 0.32 $\pm$ 0.03 \\
0402$+$265 & LHS 1625a &   DC12.7  & 0.24 $\pm$ 0.05 \\
0452$+$103 & G83-43 &  DA6.8   & 0.37 $\pm$ 0.04 \\
0541$+$260 & LSR J0544$+$2603 &   DC10.0  & 0.31 $\pm$ 0.02 \\
0546$+$234 & LSR J0549$+$2329 &  D:10.6  &  0.24 $\pm$ 0.02 \\
0922$+$005 & LHS 2139 &   D:12.2  & 0.34  $\pm$ 0.02 \\
0944$+$452B  &        &  DC16.7  & 0.13  $\pm$ 0.02 \\
1059$+$415 & SDSS  J1102$+$4113 &  DC14.9  & 0.35  $\pm$ 0.03 \\
1335$+$002 &   &  DC18.2  & 0.31 $\pm$ 0.02 \\
1632$+$177 & PG  1632$+$177 &   DA5.2   & 0.24  $\pm$ 0.01 \\
2058$+$342B & GD 392B & DC15.5 & 0.18 $\pm$ 0.03 \\
\enddata 
\tablenotetext{a}{If a single white dwarf.}
\end{deluxetable}

\subsection{Age and Kinematics}

Figure 13 shows that most of the stars in the sample are younger than 10~Gyr as would be expected for a relatively local disk sample. \citet{Gagne2018} has shown that one object in our sample, 0346$-$011, is a member of the AB Doradus moving group.  Gagn{\'e} et al. used the {\it Gaia} DR2 parallax and proper motion,
MESA Isochrones and Stellar Tracks (MIST) isochrones \citep{Choi2016}, and the same  atmosphere models and initial-final mass relation used here, to demonstrate that this massive DA1.2 star is a member of the young AB Dor group. Our analysis using the NOFS parallax and proper motion and PARSEC isochrones supports this conclusion --- we find a total age for 0346$-$011 of $115^{+125}_{-29}$~Myr, consistent with the AB Dor age given by Gagn{\'e} et al. of  ${133}^{+15}_{-20}$~Myr.

Figure 14 is a plot of the tangential velocity ($V_{\rm tan}$) of the sample as a function of total age. 
%As would be expected, there is an overall trend of increasing velocity with increasing age. 
Using the Besancon Galaxy model 
\citep{Robin2003}, \citet{Dupuy2012} calculate probabilities of population membership as a function of $V_{\rm tan}$. They find a 90\% probability of thick disk membership for  $150 \lesssim V_{\rm tan}$~km~s$^{-1} \lesssim 210$ and a $\gtrsim 50$\% probability of halo membership for $V_{\rm tan} >250 $~km~s$^{-1}$. \citet{Kilic2017} examine the faint end of the luminosity functions of samples of WDs and derive ages of 7.5$\pm$1.2~Gyr for the thin disk, and 9.3$\pm$0.6~Gyr for the thick disk. The halo age is less constrained but Kilic et al. estimate it to lie in the range  9 -- 14~Gyr.

\begin{figure}[!t]
\begin{center}
    \includegraphics[angle=-90,width=0.84\textwidth]{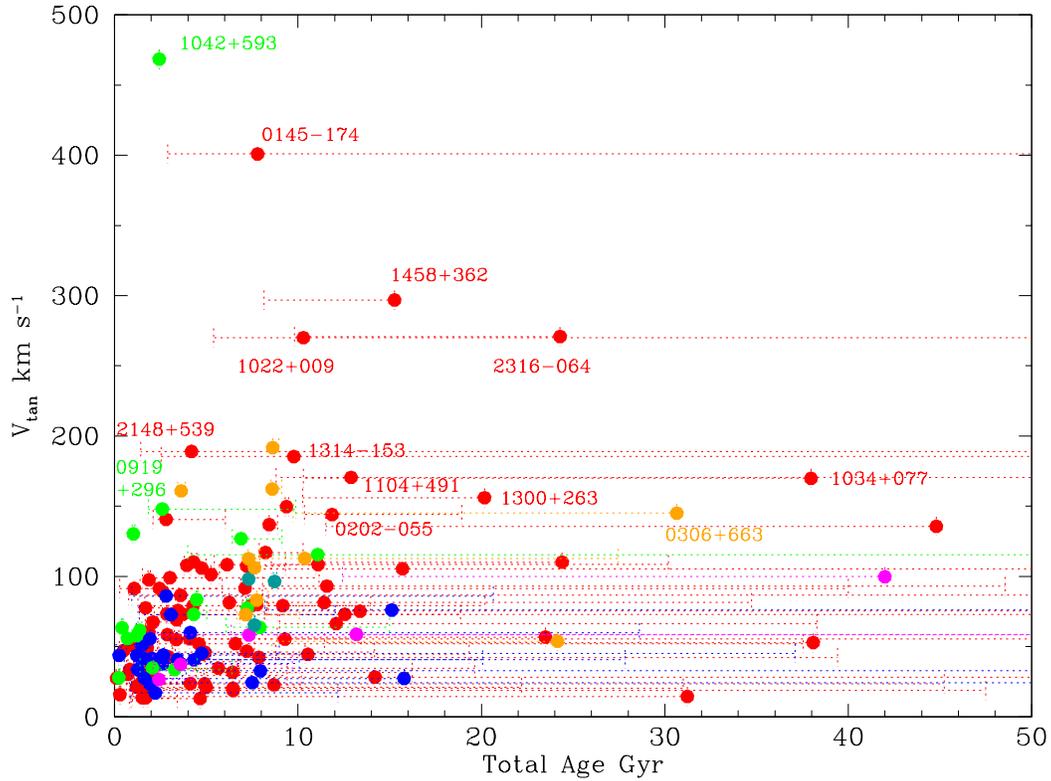}\\
\vskip -0.25in
\caption{The tangential velocity  as a function of total age. Symbol colors are as in Figure 9. Uncertainties in age are shown as dotted lines.
}
\end{center}
\end{figure}

\begin{deluxetable}{lccrrr}[!b]
\tabletypesize{\small}
\tablewidth{0pt}
\tablecaption{Candidate Halo White dwarfs}
\tablehead{
\colhead{WD Number} & \colhead{Name} & 
\colhead{Spectral}  & \colhead{Mass} & \colhead{Lower Age} & \colhead{$V_{\rm tan}$}  \\
\colhead{ } & \colhead{ } & 
\colhead{Type} & \colhead{$M_{\odot}$} & \colhead{Limit Gyr} & \colhead{km~s$^{-1}$} 
}
\startdata
0145$-$174      &       EGGR 467        &   DA6.8 & 0.55 & 2.9 & 401 \\
0202$-$055 &  & DC12.2  & 0.60 & 10.4 & 144   \\
0306$+$663 & LSR J0310$+$6634 & DC10.7  & 0.51 & 9.7 & 145 \\
0919$+$296 & LP  313-49      & DQ6.2   &   0.65 & 1.8 & 148 \\
1022$+$009 &  LP  610-10      & DA9.3 & 0.55 & 5.4 & 270 \\
1034$+$077 &  LHS  2288 & DC11.7  & 0.50 & 10.3 & 170  \\
1104$+$491 & LSR  J1107$+$4855 & DC10.8  & 0.55 & 8.8 & 171  \\
1300$+$263 &  LHS  2673 & DC11.7  & 0.53 & 10.2 & 156  \\
1314$-$153 &   LP  737-47 &  DA3.4 & 0.54 & 2.5 & 185 \\
1458$+$362 &  [MFL2000]J1500$+$3600 &  DC:10.3 & 0.54 & 8.2 & 297  \\
2148$+$539  &   G232-38 &  DA4.9 & 0.58 & 1.4 & 189 \\
2316$-$064 &  LHS  542 & DC11.2  & 0.51 & 9.8 & 271  
\enddata 
\end{deluxetable}

Table 17 lists the twelve WDs in our core sample of 179 for which  $V_{\rm tan} > 140$~km~s$^{-1}$ and the total age range encompasses 
values  $>$~9.5~Gyr. 
One of these WDs (0202$-$055) is a normal-mass cool WD with a cooling age of 9 Gyr.
Seven of them (0306$+$663, 1022$+$009, 1034$+$077, 1104$+$491, 1300$+$263, 1458$+$362, 2316$-$064)
are cool low-mass WDs which have cooling ages of a few Gyr and low-mass progenitors that spent spent several Gyr on the main sequence.  
Four others (0145$-$174, 0919$+$296, 1314$-$153, 2148$+$539)
are  not particularly cool but the possible fits include masses low enough that the progenitor time on the main sequence is high. 
Based on the Galaxy model described above, the four WDs with  $V_{\rm tan} > 250$~km~s$^{-1}$ (0145$-$174, 1022$+$009, 1458$+$362, and 2316$-$064)
have a high probability of halo membership while the remaining eight are likely to be members of the thick disk.

Total age was only determined for WDs with mass $\geq 0.5$~$M_{\odot}$ because of the likelihood that lower mass WDs are the product of binary evolution. There is one very low-temperature WD  with a mass   0.48~$M_{\odot}$ which has the largest cooling age in the sample. The DC17.4 1401$+$457 has a cooling age of 9.2 Gyr. The $V_{\rm tan}$ is low at 43~km~s$^{-1}$, however if indeed this is a single low mass WD then the main sequence lifetime would be significant, making this star very old. 

The DQ5.2 1042$+$593 has the highest velocity in our core sample ($>$~450~km~s$^{-1}$, Figure 14), but a well-constrained young-disk-like age of 2.4~Gyr.  
The astrometric measurements are robust as this WD was observed independently with two different NOFS cameras (Table 1).
This WD appears to be a less extreme version of the binary system WD 2234+222 (LP 400-22)  for which  $V_{\rm tan} = 1190$~km~s$^{-1}$ (Table 1, Table 5). \citet{Kilic2013} suggest that a dynamical interaction with other binary stars or a central black hole in a dense cluster could explain the origin of WD 2234+222. \citet{Hansen2003} has suggested that high velocity white dwarfs may be the result of 
disruption of a close binary orbit by a Type Ia supernova explosion. In the case of the massive DQ 1042$+$593 it does appear that a dynamical kick of some kind is required to explain its high velocity.

\begin{deluxetable}{lccrrrrr}[b]
%\movetabledown=1.25in
%\def\arraystretch{0.8} 
%\tabletypesize{\small}
\tabletypesize{\scriptsize}
%\tablecolumns{9}
\tablewidth{0pt}
%\rotate
\tablecaption{Ages of Resolved Double Degenerate Binaries}
\tablehead{
\colhead{WD}   & \colhead{Composition} & \colhead{$T_{\rm eff}$} &
\colhead{Mass}    & \colhead{Initial}     &  \colhead{Cooling} & \colhead{Total} & \colhead{$V_{\rm tan}$}\\
\colhead{Number} & \colhead{} & \colhead{K} &
\colhead{$M_{\odot}$} & \colhead{Mass $M_{\odot}$} & \colhead{Age Gyr}  & \colhead{Age Gyr}   & \colhead{km s$^{-1}$} 
}
\startdata
0102$+$210A  &   H & 5278 $\pm$ 89 & 0.78 $\pm$ 0.04  &  3.4 $\pm$ 0.5 & 7.10 $\pm$ 0.48 &  7.4$^{+0.3}_{-0.3}$ & 78.7\\
0102$+$210B   &  H & 4818 $\pm$ 78 & 0.62 $\pm$ 0.04  &  1.8 $\pm$ 0.7 &  7.16 $\pm$ 0.51 &  9.2$^{+4.8}_{-0.8}$ & 79.3   \\
  &   &   & & & \\
0114$-$049A   &  H & 4885 $\pm$ 75 &  0.55 $\pm$ 0.04   & 1.1 $\pm$ 0.7  &   5.65 $\pm$ 0.60  & 12.6$^{+42}_{-4.2}$ & 73.8 \\
0114$-$049B  &  He & 4979 $\pm$ 54 & 0.54  $\pm$ 0.03   & 1.0 $\pm$ 0.5  &   5.25 $\pm$ 0.31 & 15.1$^{+65}_{-6.9}$ & 74.7 \\
  &   &   & & & \\
0507$+$045A  &  H & 10884 $\pm$ 418 & 0.58 $\pm$ 0.04  &  1.4 $\pm$ 0.6 & 0.46 $\pm$ 0.05 & 4.1$^{+12.4}_{-2.4}$ & 23.9\\
0507$+$045B  &  H & 19285 $\pm$ 1872 & 0.57 $\pm$ 0.05  &  1.3 $\pm$ 0.8 & 0.06 $\pm$ 0.03 & 4.9$^{+40}_{-3.7}$ & 23.2\\
  &   &   & & & \\
0944$+$452A  &  H & 4877 $\pm$ 83 &  0.67 $\pm$ 0.04  &  2.2 $\pm$ 0.8 &  7.59 $\pm$ 0.45 & 8.7$^{+1.5}_{-0.27}$ & 22.5 \\
0944$+$452B  &  N(H)/N(He)$=10^{-2.92}$ & 3009 $\pm$ 27 & 0.13 $\pm$ 0.02  &  \nodata   &  3.07 $\pm$ 0.33 &  \nodata & 22.8 \\
  &   &   & & & \\
1012$+$083A  &  H & 6650 $\pm$ 117 &  0.59 $\pm$ 0.02  &  1.5 $\pm$ 0.5 &   1.74 $\pm$ 0.10 & 4.8$^{+5.6}_{-1.6}$ & 44.7 \\
1012$+$083B   & H & 4676 $\pm$ 70 &  0.59 $\pm$ 0.03  &  1.5 $\pm$ 0.5 &   7.12 $\pm$  0.40 & 10.5$^{+8.8}_{-2.7}$ & 45.0\\
  &   &   &  && \\
1944$+$467A   &  H & 5389 $\pm$ 93 & 0.67 $\pm$ 0.03  &  2.3 $\pm$ 0.7 &   5.23 $\pm$ 0.66 & 6.3$^{+0.7}$ & 81.7\\
1944$+$467B   &  H & 4906 $\pm$ 78 & 0.56 $\pm$ 0.03  &  1.2 $\pm$ 0.6 &   5.74 $\pm$ 0.61 & 11.4$^{+23}_{-3.4}$ & 81.6\\
  &   &   & & & \\
2058$+$342A & He & 11064 $\pm$ 490 & 0.64 $\pm$ 0.04  & 2.0 $\pm$ 0.7 &  0.55 $\pm$ 0.08 &  2.1$^{+3.2}_{-0.8}$ & 41.6\\
2058$+$342B & H & 3249 $\pm$ 75 & 0.18 $\pm$ 0.03 & \nodata        &  3.49 $\pm$ 0.43 &  \nodata              & 39.9\\
  &   &   & & & \\
2139$+$132A  &  H &  7623 $\pm$ 204 & 0.63 $\pm$ 0.04  &  1.8 $\pm$ 0.6 &   1.31 $\pm$ 0.11 & 3.1$^{+3.7}_{-1.0}$ & 73.4\\
2139$+$132B  &  unconstrained & 5301 $\pm$ 87 &  0.65 $\pm$ 0.03  &  2.0 $\pm$ 0.7 &   5.49 $\pm$ 0.52 & 7.2$^{+2.9}_{-0.5}$ & 73.7\\ 
\enddata 
\end{deluxetable}

\subsection{Ages of Resolved Double Degenerate Binaries}

Table 18 lists the parameters for the eight resolved WD binaries in the sample. Such binaries would be expected to have the same total age, within the uncertainties. The ages can therefore be used to test the robustness of our analysis. Six systems consist of normal-mass WD pairs; four of these have component ages which agree within $1.0~\sigma$ and the other two agree to 1.5 -- 2.1~$\sigma$, validating the results presented here. 

Two systems in Table 18 contain an unusually low-mass WD, and for these the primaries may provide insight into the true nature of these objects. Both low-mass components have colors suggesting they are cool. If they are in fact normal-mass unresolved binaries then the cooling time is around 10~Gyr. This could work for the 0944$+$452 system where the primary is also cool, but is unlikely to be true for the 
2058$+$342 system where the primary is warm and the total age is  $<$ 5.3~Gyr. Further exploration is postponed to future work.

The binaries may also provide constraints on the chemical evolution of WDs. For example, the 0114$-$049 system apparently consists of a nearly identical pair of objects, but the chemical composition of their atmospheres is different. Again, further work is postponed to subsequent papers.

%\clearpage

%\section{Objects of Particular Interest}

%\subsubsection{The Six-Tenths Sample}

%This work presents new parallaxes for WDs that fall into the six-tenths sample that has been used in the past to determine the WD luminosity function (Harris et al. 2001, Liebert et al. 1988). New WDs of all proper motions have been discovered since 1988, primarily faint stars and stars near the Galactic plane missed in earlier surveys.  The sample is defined as  confirmed WDs with proper motions $\ge 0\farcs 6$ yr$^{-1}$, North of $\delta = -20\deg$, and with $M_V \ge 12$. The Appendix Table A.6
%lists the 111 WDs that are now defined as the six-tenths sample, an increase over the 91 given in Harris et al. 2001. Note that some of the stars have a proper motion revised to be slightly less than $0\farcs$6 yr$^{-1}$ but they are retained in the sample --- presumably some stars that were excluded might have proper motions that should place them in the sample.

%\clearpage
\vskip 0.5in
\section{Conclusion}

Trigonometric  parallaxes have been measured at the 
Naval Observatory Flagstaff Station (NOFS) telescopes for more than fifty years. In the last thirty years, refereed  papers  that have used NOFS parallaxes to study dwarf stars  have been cited more than 5000 times. The 
Northern hemisphere  observations for the  US Naval Observatory CCD Astrograph Catalog (UCAC) were obtained at the
Flagstaff Station. The UCAC in turn provided the fundamental reference frame for the Sloan Digital Sky Survey \citep{Pier2003,Zacharias2004}. The Flagstaff Station is also where  photographic plates were scanned to produce the USNO-B catalog \citep{Monet2003}. The work done by the small group of staff at the NOFS has been of the highest quality, and has been of fundamental importance to astronomy.
The parallax work at the NOFS is now winding down and this paper will be one of the last presenting NOFS parallaxes. In the next five years the {\it Gaia} satellite \citep{Brown2018} will become the primary source of stellar parallaxes and proper motions, \href{http://www.cosmos.esa.int/web/gaia/release}{with a final release planned in 2022}.

In this paper we have presented new NOFS trigonometric parallaxes for 214 stars (or unresolved systems) at distances between 25~pc and 1.2~kpc. The observations of each star took place over periods of 3 -- 10 years, and the uncertainty in the parallax ranges from 0.2~mas to 1.6~mas with an average of 0.6~mas. Ninety percent of our sample also has a {\it Gaia} parallax, although 40\% of those objects  are flagged in the {\it Gaia} Data Release 2 as having a poor astrometric fit or excess noise.   We find generally good agreement between the NOFS and  {\it Gaia} measurements with an indication of an offset in the {\it Gaia} values of $-0.13$~mas, similar to that found by \cite{Stassun2018}. 
Eighteen stars with good quality NOFS and  {\it Gaia} measurements have parallaxes that differ significantly, and these differences
can be explained by binarity and/or by small $\sim 10$\% underestimates in the uncertainties for both sets of measurements.
% Overall   {\it Gaia} demonstrates that the NOFS measurements presented here are reliable.

More than 80\% of the stars with new NOFS parallaxes presented here are WDs and we analyse these objects in detail. 
We combine the parallaxes with photometry to determine flux-calibrated SEDs for each star. In many cases the photometry spans
a wide wavelength range from the ultraviolet to the mid-infrared, and precise fluxes are provided by the sky surveys Pan-STARRS, SDSS, UHS, UKIDSS, VISTA and WISE.
We compare the SEDs, and spectroscopy where available, to flux distributions generated by model atmospheres. In the last few years white dwarf models have improved significantly. Treatments of the non-ideal equation of state and of line broadening have advanced, and opacities are more complete  \citep{Bergeron2011,Dufour2005,Dufour2007,Kowalski2006,Tremblay2009}.  We show here that the models do an excellent job of reproducing the entire SED for this inhomogeneous sample of WDs.

The sample contains hot and cool WDs with high and low gravities and a range of chemical composition. More than 60\% of the sample have pure-hydrogen atmospheres but pure-helium, mixed hydrogen-helium and metal-rich atmospheres are also present and are generally modelled successfully (Appendix Figure 15). The highest mass object in the sample is 2349$-$031 with a mass of 1.33~$M_{\odot}$, close to the Chandrasekhar limit. Twelve WDs have extremely low masses of $< 0.4~M_{\odot}$ --- these may be the single product of binary evolution  or may themselves be unresolved binaries; we identify these candidate binary systems. 
Thirty objects in the sample have a total age of 10~Gyr or more; we identify candidate thick disk and halo WDs based on total age and tangential velocity. 
The youngest object in the sample for which an age can be determined is 0346$-$011, with an age of 115~Myr, and which \citet{Gagne2018} has shown to be a member of the AB Dor moving group. The young 0346$-$011 is also the warmest star in the sample with $T_{\rm eff} \approx 41000$~K. The five coldest stars have  $2800 \lesssim T_{\rm eff}$~K $\lesssim 3400$. For two of these the model fits give extremely low masses of 0.13 -- 0.18~$M_{\odot}$. The other three cold WDs are low-mass to a lesser extent: the model fits give masses of 0.3 -- 0.5$~M_{\odot}$.

The future holds the promise of exquisitely detailed studies of the end points of stellar evolution, the WDs. New WDs will be identified in photometric or astrometric studies; accurate distances will be available as well as precise photometry. The models will reproduce observations across a wide wavelength range, allowing the confident exploration of the physics and chemistry of these high pressure atmospheres, illuminating the evolution of these intriguing objects.

%\section{Photometry}
%\subsection{SDSS}
%\subsection{Near-Infrared}
%\subsection{Mid-Infrared}
%\section{Model Fit Figures}

%\clearpage
\acknowledgements

This research has made use of the SIMBAD database, operated at CDS, Strasbourg, France.

This publication
makes use of data from  the Pan-STARRS1 Surveys (PS1). PS1 and the PS1 public science archive have been made possible through contributions by the Institute for Astronomy, the University of Hawaii, the Pan-STARRS Project Office, the Max-Planck Society and its participating institutes, the Max Planck Institute for Astronomy, Heidelberg and the Max Planck Institute for Extraterrestrial Physics, Garching, The Johns Hopkins University, Durham University, the University of Edinburgh, the Queen's University Belfast, the Harvard-Smithsonian Center for Astrophysics, the Las Cumbres Observatory Global Telescope Network Incorporated, the National Central University of Taiwan, the Space Telescope Science Institute, the National Aeronautics and Space Administration under Grant No. NNX08AR22G issued through the Planetary Science Division of the NASA Science Mission Directorate, the National Science Foundation Grant No. AST-1238877, the University of Maryland, Eotvos Lorand University (ELTE), the Los Alamos National Laboratory, and the Gordon and Betty Moore Foundation.

This publication makes use of data from the Sloan Digital Sky Survey (SDSS).
Funding for the SDSS has been provided by the Alfred P. Sloan Foundation, the U.S. Department of Energy Office of Science, and the Participating Institutions. SDSS-IV acknowledges
support and resources from the Center for High-Performance Computing at
the University of Utah. The SDSS web site is www.sdss.org.
SDSS-IV is managed by the Astrophysical Research Consortium for the 
Participating Institutions of the SDSS Collaboration including the 
Brazilian Participation Group, the Carnegie Institution for Science, 
Carnegie Mellon University, the Chilean Participation Group, the French Participation Group, Harvard-Smithsonian Center for Astrophysics, 
Instituto de Astrof\'isica de Canarias, The Johns Hopkins University, 
Kavli Institute for the Physics and Mathematics of the Universe (IPMU) / 
University of Tokyo, the Korean Participation Group, Lawrence Berkeley National Laboratory, 
Leibniz Institut f\"ur Astrophysik Potsdam (AIP),  
Max-Planck-Institut f\"ur Astronomie (MPIA Heidelberg), 
Max-Planck-Institut f\"ur Astrophysik (MPA Garching), 
Max-Planck-Institut f\"ur Extraterrestrische Physik (MPE), 
National Astronomical Observatories of China, New Mexico State University, 
New York University, University of Notre Dame, 
Observat\'ario Nacional / MCTI, The Ohio State University, 
Pennsylvania State University, Shanghai Astronomical Observatory, 
United Kingdom Participation Group,
Universidad Nacional Aut\'onoma de M\'exico, University of Arizona, 
University of Colorado Boulder, University of Oxford, University of Portsmouth, 
University of Utah, University of Virginia, University of Washington, University of Wisconsin, 
Vanderbilt University, and Yale University.

This work is based in part on observations obtained for the VISTA Hemisphere Survey, ESO Progam, 179.A-2010 (PI: McMahon).

This work is based in part on data obtained as part of the UKIRT Infrared Deep Sky Survey and the UKIRT Hemisphere Survey (UHS). The UHS is a partnership between the UK STFC, The University of Hawaii, The University of Arizona, Lockheed Martin and NASA.

This publication makes use of data products from the Two Micron All Sky Survey, 
a joint project of the University of Massachusetts and the Infrared
Processing and Analysis Center/California Institute of Technology,
funded by the National Aeronautics and Space Administration and the
National Science Foundation.

This publication makes use of data from the Wide-field Infrared Survey Explorer, a joint project of the University of California, Los Angeles, and the Jet Propulsion Laboratory/California Institute of Technology, funded by the National Aeronautics and Space Administration.

This work is based in part on archival data obtained with the Spitzer Space Telescope, operated by the Jet Propulsion Laboratory, California Institute of Technology under a contract with NASA. 

This work is based in part on observations obtained at the Gemini Observatory and processed using the Gemini IRAF package. Gemini is operated by the Association of Universities for Research in Astronomy, Inc., under a cooperative agreement with the NSF on behalf of the Gemini partnership: the National Science Foundation (United States), the National Research Council (Canada), CONICYT (Chile), Ministerio de Ciencia, Tecnolog\'{i}a e Innovaci\'{o}n Productiva (Argentina), and Minist\'{e}rio da Ci\^{e}ncia, Tecnologia e Inova\c{c}\~{a}o (Brazil). 

This work was supported in part by the NSERC Canada
and by the Fund FQRNT (Qu{\'e}bec).

%\clearpage
\bibliography{ms_September}
\bibliographystyle{aasjournal}

\appendix

%\clearpage
% uncomment for entire table
%\input{Table19.tex}
%
%\clearpage
\begin{longrotatetable}
\begin{deluxetable}{
l
r@{ $\pm$ }r
r@{ $\pm$ }r
r@{ $\pm$ }r
r@{ $\pm$ }r
r@{ $\pm$ }r
r
r
r
r@{ $\pm$ }r}
\tabletypesize{\tiny}
\tablewidth{0pt}
\tablecaption{GAIA Astrometry and Photometry}
\tablehead{
\colhead{WD Number} &  \multicolumn{2}{c}{RA(2015)} &  \multicolumn{2}{c}{Dec(2015)} &   \multicolumn{2}{c}{$\pi$} 
 &   \multicolumn{2}{c}{$\mu$ RA}  &   \multicolumn{2}{c}{$\mu$ Dec} 
& \colhead{Goodness} & \multicolumn{2}{c}{Excess Noise}
&   \multicolumn{2}{c}{{\it Gaia} G} \\
\colhead{or Name} &   \colhead{degrees} &  \colhead{mas} &   \colhead{degrees} &  \colhead{mas}   
& \multicolumn{2}{c}{mas}
& \multicolumn{4}{c}{mas yr$^{-1}$}
& \colhead{of Fit} & \colhead{mas}   & \colhead{Significance}
&  \multicolumn{2}{c}{magnitudes}  
}
\startdata
2359$+$636\tablenotemark{a}     &       0.6033724108    &       0.057   &       63.9627280595   &       0.061   &       38.080  &       0.079   &       918.853 &       0.126   &       108.269 &       0.113   &       0.765395        &       0.089375        &       0.220290        &       17.0238 &       0.0008  \\
0002$+$729\tablenotemark{b}      &       1.2812254420    &       0.032   &       73.2197988339   &       0.028   &       28.841  &       0.033   &       188.090 &       0.064   &       122.007 &       0.053   &       10.121792       &       0.254077        &       12.341704       &       14.2866 &       0.0004  \\
0003$-$103        &       1.4832881447    &       0.173   &       $-$10.0369996049  &       0.097   &       6.496   &       0.213   &       83.637  &       0.382   &       22.531  &       0.199   &       1.867083        &       0.326108        &       0.990932        &       17.9770 &       0.0039  \\
0003$+$176      &       1.5968322482    &       0.091   &       18.0023371153   &       0.082   &       17.259  &       0.114   &       278.227 &       0.210   &       $-$322.470        &       0.156   &       $-$1.254582       &       0.000000        &       0.000000        &       16.9803 &       0.0011  \\
0042$-$064        &       11.2768904848   &       0.135   &       $-$6.1417281420   &       0.104   &       33.336  &       0.163   &       113.466 &       0.315   &       $-$684.933        &       0.197   &       $-$0.699422       &       0.000000        &       0.000000        &       17.9893 &       0.0011  \\
\enddata 
\tablecomments{Table 19 is published in its entirety in the machine-readable format.  A portion is shown here for guidance regarding its form and content.}
\vskip-0.05in
\tablenotetext{a}{NOFS and {\it Gaia} parallax values differ by more than $2 \sigma$.}
\tablenotetext{b}{The {\it Gaia} astrometric fit may be poor as indicated by a goodness of fit statistic $>3$ or the significance of excess noise being $>2$  
(\href{http://www.cosmos.esa.int/archive/documentation/GDR2/index.html}{Gaia DR2 documentation}).} 
\end{deluxetable}
\end{longrotatetable}
%\clearpage
% uncomment for entire table
%\input{Table20.tex}
%
%\clearpage
\startlongtable
%\begin{longrotatetable}
%\setlength{\tabcolsep}{3pt}
\begin{deluxetable}{lr@{ $\pm$ }rr@{ $\pm$ }rr@{ $\pm$ }rr@{ $\pm$ }rr@{ $\pm$ }r}
%\movetabledown=1.25in
%\def\arraystretch{0.8} 
%\tabletypesize{\tiny}
%\tabletypesize{\small}
%\tabletypesize{\scriptsize}
\tabletypesize{\footnotesize}
%\tablecolumns{9}
\tablewidth{0pt}
%\rotate
\tablecaption{SDSS Photometry, AB magnitudes}
\tablehead{
\colhead{WD/Name} & 
\multicolumn{2}{c}{$u$} & 
\multicolumn{2}{c}{$g$} &
\multicolumn{2}{c}{$r$} &
\multicolumn{2}{c}{$i$} &
\multicolumn{2}{c}{$z$}  
}
\startdata
0003$-$103 &      17.346 &      0.014 &      17.701 &      0.019 &      18.113 &      0.019 &      18.480 &      0.024 &      18.782 &      0.055 \\
0003$+$177 &      17.412 &      0.029 &      17.113 &      0.025 &      17.001 &      0.021 &      17.055 &      0.027 &      17.159 &      0.025 \\
0015$+$004 &      16.830 &      0.012 &      16.935 &      0.019 &      17.191 &      0.009 &      17.430 &      0.015 &      17.658 &      0.022 \\
0042$-$064 &      20.484 &      0.061 &      18.673 &      0.043 &      17.938 &      0.020 &      17.630 &      0.017 &      17.556 &      0.019 \\
0102$+$210A &      19.600 &      0.043 &      18.291 &      0.016 &      17.683 &      0.022 &      17.443 &      0.018 &      17.367 &      0.024 \\
\enddata 
\tablecomments{Table 20 is published in its entirety in the machine-readable format.  A portion is shown here for guidance regarding its form and content.}
\end{deluxetable}
%\end{longrotatetable}

%\clearpage
% uncomment for entire table
%\input{Table21.tex}
%
%\clearpage
\startlongtable
%\begin{longrotatetable}
%\setlength{\tabcolsep}{3pt}
\begin{deluxetable}{lr@{ $\pm$ }rr@{ $\pm$ }rr@{ $\pm$ }r}
%\movetabledown=1.25in
%\def\arraystretch{0.8} 
%\tabletypesize{\tiny}
%\tabletypesize{\small}
%\tabletypesize{\scriptsize}
\tabletypesize{\footnotesize}
%\tablecolumns{9}
\tablewidth{0pt}
%\rotate
\tablecaption{2MASS Photometry 2MASS System, Vega magnitudes}
\tablehead{
\colhead{WD/Name} & 
\multicolumn{2}{c}{$J$} & 
\multicolumn{2}{c}{$H$} &
\multicolumn{2}{c}{$K_s$}
}
\startdata
2359$+$636 & 15.799 & 0.067 & 15.575 & 0.136 & 15.513 & 0.216  \\
0002$+$729 & 14.615 & 0.036 & 14.597 & 0.055 & 14.758 & 0.100  \\
0003$+$177 & 16.208 & 0.126 & \nodata & \nodata & \nodata & \nodata \\
0042$-$064 & 16.691 & 0.117 & 16.631 & 0.225 & 15.682 & 0.215  \\
0102$+$210A & 16.518 & 0.097 & 16.504 & 0.198 & \nodata & \nodata \\
\enddata 
\tablecomments{Table 21 is published in its entirety in the machine-readable format.  A portion is shown here for guidance regarding its form and content.}
\end{deluxetable}

\clearpage
\startlongtable
%\begin{longrotatetable}
%\setlength{\tabcolsep}{3pt}
\begin{deluxetable}{lr@{ $\pm$ }rr@{ $\pm$ }rr@{ $\pm$ }rr@{ $\pm$ }rr@{ $\pm$ }rl}
%\movetabledown=1.25in
%\def\arraystretch{0.8} 
%\tabletypesize{\tiny}
%\tabletypesize{\small}
%\tabletypesize{\scriptsize}
\tabletypesize{\footnotesize}
%\tablecolumns{9}
\tablewidth{0pt}
%\rotate
\tablecaption{UKIDSS Photometry MKO System, Vega magnitudes}
\tablehead{
\colhead{WD/Name} & 
\multicolumn{2}{c}{$Z$} & 
\multicolumn{2}{c}{$Y$} &
\multicolumn{2}{c}{$J$} &
\multicolumn{2}{c}{$H$} &
\multicolumn{2}{c}{$K$}    & \colhead{Survey} 
}
\startdata
0015$+$004 & \nodata & \nodata & 17.353 & 0.015 & 17.346 & 0.017 & 17.376 & 0.066 & 17.554 & 0.114 & LAS10  \\
0112$+$104  & \nodata & \nodata & 15.860 & 0.007 & 15.945 & 0.010 & 16.027 & 0.018 & 16.059 & 0.029 & LAS10  \\
0135$-$005 & \nodata & \nodata & 13.662 & 0.002 & 12.985 & 0.001 & 12.585 & 0.002 & 12.360 & 0.002 & LAS10  \\
0235$+$064 & \nodata & \nodata & 15.759 & 0.007 & \nodata & \nodata & 15.794 & 0.013 & 15.856 & 0.023 & LAS10  \\
PG 0235$+$064B & \nodata & \nodata & 12.209 & 0.001 & \nodata & \nodata & 11.447 & 0.001 & 10.997 & 0.001 & LAS10  \\
0300$-$013 & \nodata & \nodata & 15.781 & 0.006 & 15.821 & 0.010 & 15.859 & 0.011 & 15.789 & 0.017 & LAS10  \\
0339$+$523 & \nodata & \nodata & \nodata & \nodata & 15.979 & 0.005 & 15.997 & 0.011 & 16.046 & 0.025 & GPS10 \\
0343$+$247 & 17.374 & 0.015 & 17.372 & 0.016 & 17.601 & 0.031 & 17.919 & 0.062 & 18.982 & 0.238 & GCS10 \\
\enddata 
\tablecomments{``apermag3'' values are given as appropriate for point sources; DXS --- Deep eXtragalactic Survey, GCS --- 
Galactic Clusters Survey, GPS --- Galactic Plane Survey, LAS --- Large Area Survey.}
\end{deluxetable}

\begin{deluxetable}{lr@{ $\pm$ }rr@{ $\pm$ }rr@{ $\pm$ }rr@{ $\pm$ }rr@{ $\pm$ }rl}
%\movetabledown=1.25in
%\def\arraystretch{0.8} 
\tabletypesize{\small}
%\tabletypesize{\scriptsize}
%\tablecolumns{9}
\tablewidth{0pt}
%\rotate
\tablecaption{VISTA Photometry VISTA System, Vega magnitudes}
\tablehead{
\colhead{WD/Name} & 
\multicolumn{2}{c}{$Z$} & 
\multicolumn{2}{c}{$Y$} &
\multicolumn{2}{c}{$J$} &
\multicolumn{2}{c}{$H$} &
\multicolumn{2}{c}{$K$}    & \colhead{Survey}
}
\startdata
0112$-$018  & \nodata & \nodata & 16.401 & 0.010 & 16.182 & 0.010 & 15.940 & 0.020 & 15.860 & 0.020 & VHS5  \\
0114$-$049A & \nodata & \nodata & \nodata & \nodata & 16.557 & 0.009 & 16.340 & 0.015 & 16.250 & 0.030 & VHS5  \\
0115$-$049B & \nodata & \nodata & \nodata & \nodata & 16.599 & 0.010 & 16.406 & 0.016 & 16.324 & 0.032 & VHS5  \\
0202$-$055 & \nodata & \nodata & \nodata & \nodata & 16.531 & 0.009 & 16.490 & 0.010 & 16.480 & 0.040 & VHS5  \\
0918$-$172 & \nodata & \nodata & \nodata & \nodata & 16.755 & 0.017 & \nodata & \nodata & 16.495 & 0.052 & VHS5  \\
LP 787$-$25 & \nodata & \nodata & \nodata & \nodata & 12.748 & 0.001 & \nodata & \nodata & 11.919 & 0.001 & VHS5  \\
1053$-$092 & \nodata & \nodata & 17.024 & 0.029 & 17.156 & 0.050 & 17.281 & 0.099 & 17.655 & 0.274 & VHS5  \\
1212$-$022 & 16.984 & 0.004 & 16.936 & 0.007 & 16.857 & 0.007 & 16.609 & 0.013 & 16.462 & 0.018 & VIKING4  \\
\enddata 
\tablecomments{``apermag3'' values are given as appropriate for point sources; VHS --- VISTA Hemisphere Survey, VIKING --- VISTA Kilo-degree Infrared Galaxy survey.}
\end{deluxetable}

\begin{deluxetable}{lr@{ $\pm$ }rlr@{ $\pm$ }rlr@{ $\pm$ }r}
\tabletypesize{\small}
\tablewidth{0pt}
\tablecaption{UKIRT Hemisphere Survey MKO System, Vega magnitudes}
\tablehead{
\colhead{WD/Name} & \multicolumn{2}{c}{$J$} &
\colhead{WD/Name} & \multicolumn{2}{c}{$J$} &
\colhead{WD/Name} & \multicolumn{2}{c}{$J$} 
}
\startdata
0003$+$177 & 16.609 & 0.018 & 0955$+$247 & 14.678 & 0.004 & 1630$+$245 & 17.641 & 0.045 \\
0102$+$210A & 16.528 & 0.014 & 1002$+$430 & 16.636 & 0.020 & 1632$+$177 & 13.016 & 0.002 \\
0102$+$210B & 16.452 & 0.013 & 1056$+$345 & 15.657 & 0.008 & 1636$+$057 & 15.963 & 0.010 \\
GD 274 & 11.305 & 0.001 & 1100$+$211 & 17.037 & 0.022 & 1645$+$325 & 14.078 & 0.003 \\
0203$+$183 & 16.651 & 0.019 & 1101$+$364 & 14.823 & 0.005 & 1658$+$440 & 15.498 & 0.007 \\
0203$+$207 & 17.940 & 0.038 & 1104$+$491 & 17.196 & 0.027 & 1704$+$481B & 14.112 & 0.003 \\
0228$+$269 & 15.797 & 0.011 & 1108$+$207 & 15.947 & 0.009 & 1704$+$481A & 14.769 & 0.005 \\
0239$+$109 & 15.739 & 0.011 & 1124$+$595 & 15.046 & 0.006 & 1709$+$230 & 15.326 & 0.007 \\
\enddata 
\tablecomments{``apermag3'' values are given as appropriate for point sources.}
\end{deluxetable}

\clearpage
% uncomment for entire table
%\input{Table24.tex}
%
%\clearpage
\startlongtable
%\begin{longrotatetable}
%\setlength{\tabcolsep}{3pt}
\begin{deluxetable}{lr@{ $\pm$ }rr@{ $\pm$ }rr@{ $\pm$ }rr@{ $\pm$ }rr@{ $\pm$ }rr@{ $\pm$ }rr@{ $\pm$ }rr@{ $\pm$ }r}
%\movetabledown=1.25in
%\def\arraystretch{0.8} 
\tabletypesize{\tiny}
%\tabletypesize{\small}
%\tabletypesize{\scriptsize}
%\tablecolumns{9}
\tablewidth{0pt}
%\rotate
\tablecaption{{\it WISE} and {\it Spitzer} Photometry, Vega magnitudes}
\tablehead{
\colhead{WD/Name} & 
\multicolumn{2}{c}{W1} & 
\multicolumn{2}{c}{W2} &
\multicolumn{2}{c}{W3} &
\multicolumn{2}{c}{W4} &
\multicolumn{2}{c}{$3.6\mu$m} &
\multicolumn{2}{c}{$4.5\mu$m}  &
\multicolumn{2}{c}{$5.8\mu$m} &
\multicolumn{2}{c}{$8.0\mu$m}
}
\startdata
2359$+$636 & 14.884 & 0.030 & 15.192 & 0.066 & \nodata & \nodata & \nodata & \nodata & \nodata & \nodata & \nodata & \nodata & \nodata & \nodata &  \nodata &   \nodata\\
0002$+$729 & 14.743 & 0.032 & 14.830 & 0.054 & \nodata & \nodata & \nodata & \nodata & \nodata & \nodata & 14.743 & 0.037 & \nodata & \nodata & 14.369 & 0.015 \\
0015$+$004 & 17.224 & 0.155 & \nodata & \nodata & \nodata & \nodata & \nodata & \nodata & 17.418 & 0.006 & 17.325 & 0.007 & \nodata & \nodata & \nodata & \nodata \\
0042$-$064 & 16.340 & 0.072 & \nodata & \nodata & \nodata & \nodata & \nodata & \nodata & \nodata & \nodata & \nodata & \nodata & \nodata & \nodata & \nodata & \nodata \\
0102$+$210A & 15.962 & 0.055 & 15.667 & 0.123 & \nodata & \nodata & \nodata & \nodata & 16.102 & 0.035 & 16.001 & 0.039 & 15.829 & 0.193 & 16.155 & 0.588 \\
\enddata 
\tablecomments{Table 25 is published in its entirety in the machine-readable format.  A portion is shown here for guidance regarding its form and content.}
\end{deluxetable}
%\end{longrotatetable}

\clearpage
\begin{figure}
\vskip -0.1in
\begin{center}
    \includegraphics[angle=0,width=0.9\textwidth]{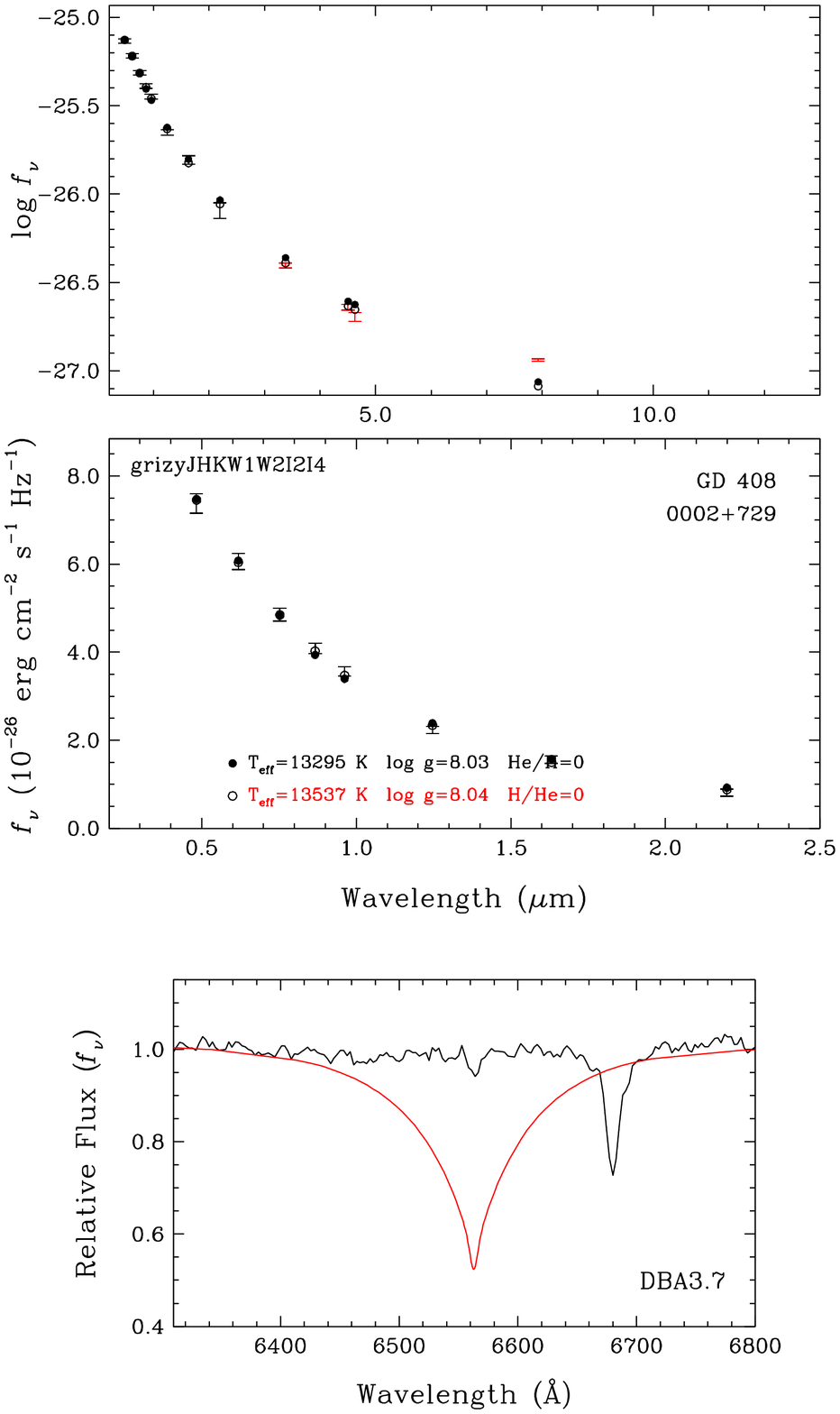}\\
\vskip -0.2in
\caption{Example Figure of the model comparison to each WD. The Figure is published in its entirety   in the on-line Journal.
}
\end{center}
\end{figure}

%\facility{USNO:61in (TEK2K, EEV24, TI800), USNO:40in}

\end{document}